\documentclass[12pt]{iopart}

\usepackage{iopams}
\usepackage{graphicx}
\usepackage{enumerate}
\usepackage{bm}

\begin{document}

\title[Quasi-2D nonlinear evolution of HMRI]{Quasi-two-dimensional nonlinear evolution of helical magnetorotational
instability in a magnetized Taylor-Couette flow}

\author{G. Mamatsashvili$^{1,2,3}$, F. Stefani$^{1}$, A. Guseva$^{4}$ and M. Avila$^{4}$}
\address{$^1$Helmholtz-Zentrum Dresden-Rossendorf, Dresden 01328, Germany\\
$^2$Department of Physics, Faculty of Exact and Natural Sciences, Tbilisi State University, Tbilisi 0179, Georgia\\
$^3$Abastumani Astrophysical Observatory, Ilia State University, Tbilisi 0162, Georgia\\
$^4$University of Bremen, Center of Applied Space Technology and
Microgravity (ZARM), Bremen 28359, Germany}

\ead{g.mamatsashvili@hzdr.de} \vspace{10pt}

\begin{abstract}
Magnetorotational instability (MRI) is one of the fundamental
processes in astrophysics, driving angular momentum transport and
mass accretion in a wide variety of cosmic objects. Despite much
theoretical/numerical and experimental efforts over the last
decades, its saturation mechanism and amplitude, which sets the
angular momentum transport rate, remains not well understood,
especially in the limit of high resistivity, or small magnetic
Prandtl numbers typical to interiors (dead zones) of protoplanetary
disks, liquid cores of planets and liquid metals in laboratory.
Using direct numerical simulations, in this paper we investigate the
nonlinear development and saturation properties of the helical
magnetorotational instability (HMRI) -- a relative of the standard
MRI -- in a magnetized Taylor-Couette flow at very low magnetic
Prandtl number (correspondingly at low magnetic Reynolds number)
relevant to liquid metals. For simplicity, the ratio of azimuthal
field to axial field is kept fixed. From the linear theory of HMRI,
it is known that the Elsasser number, or interaction parameter
determines its growth rate and plays a special role in the dynamics.
We show that this parameter is also important in the nonlinear
problem. By increasing its value, a sudden transition from weakly
nonlinear, where the system is slightly above the linear stability
threshold, to strongly nonlinear, or turbulent regime occurs. We
calculate the azimuthal and axial energy spectra corresponding to
these two regimes and show that they differ qualitatively.
Remarkably, the nonlinear state remains in all cases nearly
axisymmetric suggesting that HMRI turbulence is quasi
two-dimensional in nature. Although the contribution of
non-axisymmetric modes increases moderately with the Elsasser
number, their total energy remains much smaller than that of the
axisymmetric ones.
\end{abstract}

\maketitle

\section{Introduction}

The magnetorotational instability (MRI,
\cite{Velikhov59,Balbus_Hawley98}) is one of the most important
processes in magnetized differentially rotating conducting fluids,
which largely determines their dynamics and evolution. It is a
powerful linear instability arising as a result of the combined
effect of weak magnetic field and radially decreasing angular
velocity. The MRI is believed to operate in a vast variety of cosmic
objects, ranging from astrophysical disks and stars to liquid-metal
cores of planets. While discovered as early as 1959
\cite{Velikhov59}, the astrophysical significance of MRI was first
recognized only three decades later by Balbus \& Hawley
\cite{Balbus_Hawley91}, who demonstrated that weak magnetic fields
can destabilize Keplerian accretion disks around such diverse
objects as supermassive black holes, black holes in X-ray binaries
and young stellar objects, which would otherwise be linearly stable
according to Rayleigh's criterion. The linear growth of MRI, which
taps into the free energy of differential rotation, eventually
breaks down into magnetohydrodynamic (MHD) turbulence
\cite{Hawley_etal95,Fromang13}. This turbulence transports angular
momentum outward and, as a consequence, matter inward in the disk,
yielding mass accretion rates onto the central object close to
observationally inferred values. More recently, MRI has also been
discussed for explaining angular momentum transport in the Sun,
massive stars and neutron stars
\cite{Kagan_Wheeler14,Wheeler_etal15,Masada_etal15,Spada_etal16} and
also in the context of the geodynamo
\cite{Petitdemange_etal08,Petitdemange_etal13}.

The MRI was originally discovered theoretically in a classical
Taylor-Couette (TC) flow of a conducting fluid between two
concentric rotating cylinders threaded by an external vertical
(along the common axis of cylinders) magnetic field
\cite{Velikhov59,Chandrasekhar60} -- a setup which is also best
suited and most frequently used nowadays for experimental
investigations. The conducting medium filling the gap between
cylinders is usually a liquid metal (sodium, gallium) characterized
by an extremely small ratio of viscosity to Ohmic resistivity, or
magnetic Prandtl number, $Pm=\nu/\eta \sim 10^{-6}-10^{-5}$. By
suitably adjusting the rotation rates of outer and inner cylinders,
the TC flow profile can be made very close to the Keplerian one
\cite{Ji_etal06,Schartman_etal12,Edlund_Ji15,Lopez_Avila17},
offering a unique possibility to study the disk MRI problem in
laboratory as well, which has been up to now mostly carried out both
via analytical means and numerical simulations.

First experimental efforts to study MRI in the laboratory were made
at the University of Maryland \cite{Sisan_etal04}, at Princeton
University \cite{Nornberg_etal10,Roach_etal12}, and at
Helmholtz-Zentrum Dresden-Rossendorf (HZDR)
\cite{Stefani_etal06,Stefani_etal09,Seilmayer_etal14}. The liquid
sodium spherical Couette experiment in Maryland had produced
coherent velocity/magnetic field fluctuations showing up in a
parameter region reminiscent of MRI \cite{Sisan_etal04}. The liquid
GaInSn TC experiments in Princeton was designed to investigate the
standard version of MRI (SMRI) with only an axial magnetic field
being imposed. In this case, the azimuthal magnetic field (which is
an essential participant in the MRI process) must be produced from
the axial field by induction effects, which are proportional to the
magnetic Reynolds number $Rm$ of the flow. $Rm$, in turn, is
proportional to the hydrodynamic (HD) Reynolds number $Re$ according
to $Rm = Pm\cdot Re$. Therefore, in order to achieve $Rm \gtrsim 1$
necessary for SMRI to operate (see e.g.,
\cite{Sano_Miyama99,Ji_etal01}), taking into account the very small
values of $Pm$, $Re \sim 10^5-10^6$ is needed. This makes SMRI
experiments with TC flows extremely challenging, although evidence
for slow magneto-Coriolis waves \cite{Nornberg_etal10} and for a
free-Shercliff layer instability \cite{Roach_etal12} have already
been obtained. Within the DRESDYN project at HZDR
\cite{Stefani_etal12}, it is planned to set-up a large liquid sodium
TC experiment, which will allow  to broaden the parameter range in
such a way as to make SMRI reachable.

On the theoretical side, SMRI in TC flow has been extensively
studied both in the linear and nonlinear regimes, with more focus on
the low-$Pm$ regime as it is typical in experiments. Still, the
typical Reynolds numbers $Re\sim 10^3$ used in these analyses are
orders of magnitude smaller than experimental values $Re\gtrsim
10^6$, which are extremely demanding from the computational point of
view. The linear analysis identified critical magnetic Reynolds
numbers for the onset of the instability and discussed possibilities
for its experimental detection (e.g.,
\cite{Ji_etal01,Ruediger_Zhang01,Goodman_Ji02,Kirillov_Stefani10}).
In particular, it was shown that the linear growth rate is
determined by magnetic Reynolds number and Lundquist number (i.e.,
ratio of the magnetic diffusion time to the Alfv\'en crossing time),
which should be both larger than unity for SMRI to operate. The
saturation properties of SMRI following its exponential growth was
investigated both in the weakly nonlinear regime, i.e., near the
instability threshold,
\cite{Umurhan_etal07a,Umurhan_etal07b,Clark_Oishi16a,Clark_Oishi16b}
and in the fully nonlinear regime
\cite{Knobloch_Julien05,Liu_etal06a,Liu08,Ebrahimi_etal09,Gissinger_etal12,Gellert_etal12,Wei_etal16}.
It was demonstrated that SMRI saturates on the resistive time scale
by modifying (reducing) the background velocity shear responsible
for it and strengthening the background field. The dependence of the
saturation level on the imposed field as well as on the magnetic
Reynolds number, Prandtl number and Lundquist number was explored in
greater detail.

While SMRI has been explored quite extensively both in TC flows and
astrophysical disk context since the 1990s, its relative -- helical
magnetorotational instability (HMRI) -- has become a subject of
active theoretical and experimental research only in the last
decade. These studies were initiated by Hollerbach \& R{\"u}diger
\cite{Hollerbach_Ruediger05}, who realized that adding an azimuthal
magnetic field to the axial field can destabilize highly resistive
flows at much smaller field strengths and at several orders of
magnitude smaller Reynolds numbers, reducing the critical value from
$Re\sim 10^6$ needed for SMRI with purely axial field to $Re\sim
10^3$. Because this instability takes place in the presence of
helical magnetic field, composed of azimuthal and axial components,
it was  termed HMRI. Like the SMRI, the HMRI also draws free energy
from the background shear flow and transports angular momentum
outward. Subsequently, HMRI has been widely studied theoretically by
means of linear modal stability analysis
\cite{Liu_etal06,Priede_etal07,Priede_Gerbeth09,Kirillov_Stefani10,
Priede11,Kirillov_Stefani13,Kirillov_etal14,Ruediger_etal06}. It was
shown that HMRI represents a destabilized inertial wave
\cite{Liu_etal06} and its growth rate is governed by the Reynolds
number and the Hartmann number, $Ha$, characterizing the magnetic
field strength. As a result, HMRI can persist even at extremely
small magnetic Prandtl numbers typical to liquid metals, in contrast
to SMRI, which is generally suppressed under this condition. Another
difference with SMRI is that HMRI is restricted to rotational
profiles with comparably steep negative shear or extremely steep
positive shear, smaller than so-called the lower Liu limit or larger
than the upper Liu limit, respectively \cite{Liu_etal06,Priede11}.
Although the transition between HMRI and SMRI is monotonic when
decreasing the ratio of azimuthal to axial field
\cite{Hollerbach_Ruediger05}, the mathematics of this connection
turned out to be quite subtle, including the formation of a spectral
exceptional point where the original inertial mode coalesces with
the slow magnetocoriolis mode \cite{Kirillov_Stefani10}. Moreover,
the recent nonmodal analysis revealed a fundamental connection
between the modal growth of HMRI and the nonmodal dynamics in the
corresponding HD problem \cite{Mamatsashvili_Stefani16}. The
relevance of HMRI for astrophysical Keplerian disks has also been
considered: it seems to be a promising candidate to replace SMRI in
weakly ionized parts of disks with small $Pm$, for example in the
``dead zones'' of protoplanetary disks or the outer parts of
accretion disks around black holes. However, this issue is not yet
fully resolved and under debate, because under standard conditions
Keplerian shear alone might not be steep enough for HMRI to work
\cite{Liu_etal06} unless other physical factors, such as specific
electrical boundary conditions
\cite{Ruediger_Hollerbach07,Ruediger_Schultz08} or additional axial
currents within the fluid are involved \cite{Kirillov_Stefani13}.

The remarkable feature of HMRI to survive at very small magnetic
Prandtl numbers and to set in at moderate Reynolds numbers makes it
an ideal playground for experimental studies with liquid metals.
Indeed, shortly after its theoretical discovery, a series of
specially designed TC experiments
\cite{Stefani_etal06,Stefani_etal09} provided the first experimental
evidence of HMRI at the liquid metal facility PROMISE and reproduced
the main results of the linear theory, such as the stability
threshold, the wavenumber and frequency of the HMRI-wave. In the
experiments, the saturation amplitude as well as the propagation
speed of the HMRI wave were measured in detail as a function of the
system parameters ($Re$, $Ha$, ratio of rotation frequencies of
outer and inner cylinders, etc.) and some differences with the
numerical results were pointed out (see e.g., figure 10 in
\cite{Stefani_etal09}).

This prompted further theoretical studies of HMRI in TC flows both
with and without endcaps, focusing more on its nonlinear development
and saturation. However, even today this is still a less explored
area than the linear HMRI. First axisymmetric numerical simulations
explored the dependence of the established HMRI-wave amplitude and
propagation speed on the Reynolds number and the rotation ratio of
the inner and outer cylinders as well as the role of endcaps
\cite{Szklarski_Ruediger06,Liu_etal07,Szklarski07}. However, the
parameter range adopted in these studies was much narrower than that
in the above experiments, being near the linear instability
threshold of HMRI, when its dynamics is weakly nonlinear (see also
\cite{Clark_Oishi16b}). More recent axisymmetric simulations
\cite{Child_Hollerbach16,Hollerbach_etal17} looked into the
intrinsic nonlinear dynamics of HMRI in an axially infinite TC
domain, avoiding complications because of the endcaps
\cite{Avila12}. Different types of nonlinear regimes were analyzed
and the applicability of the generalized quasi-linear approximation
was tested by tracing the dynamics of large-scale Fourier modes.
Despite these efforts, which undoubtedly contributed to a better
understanding of the nonlinear dynamics of HMRI, detailed physics of
its saturation and sustenance is still missing, especially when
comparison with experiments is concerned. Evidently, it is this
nonlinear saturated state which is observed in experiments, while
the initial transient phase is much harder to identify due to the
limited sensitivity of the applied ultrasonic flow measurement
techniques.

In some respect, more  progress has been made recently on the
nonlinear dynamics of a ``sibling'' of HMRI -- the azimuthal
magnetorotational instability (AMRI). This non-axisymmetric
instability, which emerges in the presence of an imposed purely
azimuthal fields in TC liquid metal flows, had been first identified
after HMRI \cite{Seilmayer_etal14,Hollerbach_etal10} (see also
\cite{Kirillov_etal14} and references therein). Using numerical
simulations, Guseva et al.
\cite{Guseva_etal15,Guseva_etal16,Guseva_etal17} probed much broader
ranges of Reynolds and Hartmann numbers than those done for HMRI so
far and identified different regimes of nonlinear saturation, from
supercritical Hopf bifurcation near the linear instability threshold
up to a catastrophic transition to spatio-temporal defects, which
are mediated by a subcritical subharmonic Hopf bifurcation, and
ultimately to turbulence. The scaling of the angular momentum
transport of AMRI in the nonlinear regime with respect to Reynolds,
Hartmann and Prandtl numbers was also explored in these papers.

Motivated by the experimental results of \cite{Stefani_etal09} and
by the recent progress on understanding the nonlinear dynamics of
AMRI \cite{Guseva_etal15}, in this paper we investigate the
evolution of HMRI, from its linear exponential growth to nonlinear
saturation in liquid metal TC flows at small magnetic Prandtl
numbers using numerical simulations. We consider infinite cylinders
with periodic boundary conditions in the axial direction in order to
avoid complications because of the endcaps (Ekman pumping) and
concentrate on the intrinsic dynamics of HMRI driven by the
combination of an imposed helical magnetic field and differential
rotation of the flow. Distinctly from the above-mentioned previous
studies on nonlinear HMRI, we do not restrict ourselves to
axisymmetric perturbations only and allow non-axisymmetric modes to
naturally develop during dynamical evolution, so that we can
quantify the degree of non-axisymmetry of the saturated state of
HMRI depending on the system parameters. It is well known that HMRI
is axisymmetric in the linear regime, i.e. the most unstable modes
do not have any azimuthal variation \cite{Hollerbach_Ruediger05,
Priede_etal07,Priede11,Child_Hollerbach16}, and it is therefore
important to know if this feature is retained in the nonlinear
regime, which in turn can shed light on the saturation mechanism. We
show that different regimes of the saturation are realized in the
flow and analyze the characteristics of these states, such as
angular momentum transport and energy spectra. This study, aiming at
understanding the basic nonlinear dynamics of HMRI -- underlying
mechanism and properties of its saturation -- is intended to be
guiding and preparatory for the upcoming liquid sodium TC
experiments in the DRESDYN facility at HZDR, which will combine and
enhance the previous experiments on HMRI \cite{Stefani_etal09}, AMRI
\cite{Seilmayer_etal14} and on the current-driven kink-type Tayler
instability \cite{Seilmayer_etal12}.

The paper is organized as follows. The basic equations are
introduced in section 2. Direct numerical simulations of the
nonlinear saturation and evolution of HMRI as well as the spectral
characteristics of the nonlinear state are presented in section 3.
Summary and conclusions are given in section 4.

\section{Main equations}

The basic equations of non-ideal MHD governing the motion of an
incompressible conductive fluid with constant kinematic viscosity
$\nu$ and Ohmic resistivity $\eta$ are
\begin{equation}
\frac{\partial {\bf u}}{\partial t}+({\bf u}\cdot \nabla) {\bf
u}=-\frac{1}{\rho}\nabla p+\frac{1}{\mu_0\rho}(\nabla \times {\bf
B})\times {\bf B} + \nu\nabla^2 {\bf u},
\end{equation}
\begin{equation}
\frac{\partial {\bf B}}{\partial t}=\nabla\times \left( {\bf
u}\times {\bf B}\right)+\eta\nabla^2{\bf B},
\end{equation}
\begin{equation}
\nabla\cdot {\bf u}=0,~~~\nabla\cdot {\bf B}=0,
\end{equation}
where $\rho$ is the constant density, $p$ is the thermal pressure,
${\bf u}$ is the velocity, ${\bf B}$ is the magnetic field and
$\mu_0$ is the magnetic permeability of vacuum.

The equilibrium state is an axisymmetric cylindrical magnetized TC
flow between two coaxial cylinders with the inner radius $R_i$ and
the outer radius $R_o$, rotating with the angular velocities
$\Omega_i$ and $\Omega_o$, respectively. The flow is threaded by an
externally imposed helical magnetic field ${\bf
B}_0=(0,B_{0\phi}(r),B_{0z})$ consisting of constant axial,
$B_{0z}$, and radially varying current-free azimuthal,
$B_{0\phi}(r)=\beta B_{0z}R_i/r$, components in cylindrical
coordinates $(r,\phi,z)$, where $\beta$ is the dimensionless
parameter characterizing the helicity of the field. We ignore the
effects of endcaps in order to gain insight into the basic/intrinsic
nonlinear evolution of HMRI. The ratio of the radii $a\equiv
R_i/R_o=0.5$ and the height of the cylinders, $L_z=10d$, where
$d=R_o-R_i=R_i$ is the gap width, are chosen as in PROMISE
\cite{Stefani_etal09}. An unperturbed flow between the cylinders, as
follows from Eqs. (1)-(3), is a standard vertically uniform and
axisymmetric TC flow, ${\bf U}_0=(0,r\Omega(r),0)$, with the angular
velocity given by
\begin{equation}
\Omega(r)=\frac{\Omega_oR_o^2-\Omega_iR_i^2}{R_o^2-R_i^2}+\frac{\Omega_i-\Omega_o}{R_o^2-R_i^2}\frac{R_i^2R_o^2}{r^2},
\end{equation}
pressure $p_0(r)$ and constant density $\rho_0$, satisfying the
hydrostatic balance equation $\rho_0r\Omega^2=dp_0/dr$. The rotation
ratio of the cylinders is fixed to $\mu=\Omega_o/\Omega_i=0.27$,
slightly larger than the Rayleigh line $\mu_c=0.25$ at
$R_i/R_o=0.5$, ensuring that purely HD instabilities are excluded,
so that the flow can become unstable solely due to the presence of
the magnetic field. In the following, we introduce the
nondimensional variables by using the gap width $d$ as the unit of
length, $\Omega_i^{-1}$ as the unit of time, $\Omega_id$ as the unit
of velocity and $\rho_0\Omega_i^2d^2$ as the unit of pressure and
energy density. (Since in the present case $d=R_i$, this velocity
scale is in fact the rotational velocity of the inner cylinder,
$\Omega_iR_i$.) The magnetic field is normalized by the imposed
axial field $B_{0z}$. The Reynolds number is defined in terms of the
rotation rate of the inner cylinder
\[
Re=\frac{\Omega_id^2}{\nu}.
\]
Its value is chosen to be $Re = 6000$ throughout the paper, which is
within the range used in the related experiments
\cite{Stefani_etal09}. In this paper, we consider a highly resistive
fluid with very small magnetic Prandtl number,
$Pm=\nu/\eta=1.4\times 10^{-6}$, typical to the liquid metal alloy
GaInSn, so that the corresponding magnetic Reynolds number of the
flow is also small, $Rm=Re\cdot Pm=8.4\times 10^{-3}$.

The strength of the imposed axial field is measured by the Hartmann
number,
\[
Ha=\frac{B_{0z}d}{\sqrt{\mu_0\rho_0\nu\eta}},
\]
whereas the strength of the azimuthal field relative to the vertical
one is measured by the helicity $\beta$ parameter introduced above.
Everywhere below this parameter will be fixed to $\beta=4.53$, as
adopted in some of the experiments \cite{Stefani_etal09}. Previous
linear analysis showed that HMRI is effective for relatively strong
azimuthal fields, $\beta \gtrsim 1$ (e.g.,
\cite{Hollerbach_Ruediger05,Liu_etal06,Kirillov_Stefani10,Kirillov_etal14}).

Consider perturbations of the velocity, pressure and magnetic field
about the equilibrium, ${\bf u}'={\bf u}-{\bf U}_0$, $p'=p-p_0$,
${\bf b}'={\bf B}-{\bf B}_0$. In the limit $Rm \ll 1$, which applies
here, the magnetic field perturbation induced by the perturbed flow
is much smaller than the imposed field and scales with $Rm$
\cite{Zikanov_Thess98,Willis_Barenghi02,Liu_etal06}, so we change it
to ${\bf b}'\rightarrow Rm\cdot {\bf b}'$, such that the new ${\bf
b}'$ is comparable to the perturbed velocity ${\bf u}$. Substituting
this into Eqs. (1)-(3), taking into account that the imposed
magnetic field is current-free, and using the above normalization,
we arrive at the following equations governing perturbations with
arbitrary amplitude in nondimensional form to first order in $Rm$
(primes henceforth will be omitted):
\begin{equation}
\frac{\partial {\bf u}}{\partial t}+({\bf U}_0\cdot \nabla) {\bf
u}=-\nabla p-({\bf u}\cdot \nabla) {\bf U}_0+ \frac{Ha^2}{Re}(\nabla
\times {\bf b})\times {\bf B}_0 + \frac{1}{Re}\nabla^2 {\bf u}-({\bf
u}\cdot \nabla) {\bf u},
\end{equation}
\begin{equation}
Rm\left[\frac{\partial {\bf b}}{\partial t}-\nabla\times \left({\bf
U}_0\times {\bf b}\right)-\nabla\times \left({\bf u}\times {\bf
b}\right)\right]=\nabla\times \left({\bf u}\times {\bf
B}_0\right)+\nabla^2{\bf b},
\end{equation}
\begin{equation}
\nabla\cdot {\bf u}=0,~~~\nabla\cdot {\bf b}=0,
\end{equation}
where the terms proportional to $Rm$ on the left hand side of
induction equation (6), including the time-derivative, are usually
neglected in the limit of small $Rm\rightarrow 0$ (inductionless, or
quasi-static approximation, see e.g.,
\cite{Willis_Barenghi02,Szklarski_Ruediger06,Child_Hollerbach16}).
In that case, the advective derivative, or the inertial term
involving the perturbed velocity, $({\bf u}\cdot \nabla) {\bf u}$,
is the only nonlinearity that remains in these equations. Here we
preferred to keep these small terms for completeness. In the chosen
nondimensional units, the rotational frequency (4) of the background
velocity becomes
\[
\Omega(r)=\frac{1}{1-a^2}\left(\mu-a^2+\frac{a^2}{(1-a)^2}\frac{1-\mu}{r^2}\right)
\]
and the imposed field is ${\bf B}_0=(0,\beta/r,1)$. Since $\eta,
\mu, \beta$ and $Re$ are fixed, only $Ha$ is a free parameter,
entering Eqs. (5)-(7) through the Elsasser number, or the
interaction parameter:
\[
\Lambda\equiv \frac{Ha^2}{Re}=\frac{B_{0z}^2}{\mu_0\rho_0\eta
\Omega_i},
\]
which is composed of the imposed magnetic field, rotation and
resistivity and is independent of viscosity. The interaction
parameter has been shown to play a special role in the linear
dynamics of the HMRI, as it determines its modal growth rate, the
corresponding critical wavenumber and other characteristics
\cite{Priede11,Kirillov_etal14,Mamatsashvili_Stefani16}.
Specifically, in the inviscid case ($Re\rightarrow \infty$), the
HMRI can exist in a certain range of $\Lambda$ and disappears at
sufficiently large or small values of this parameter. In the first
limit, the magnetic field is too strong and stabilizes the flow,
whereas in the second limit it is too weak to support the
instability. At finite $Re$, this implies that for a fixed $Ha$,
HMRI operates in a range of $Re$ and for a fixed $Re$, in a range of
$Ha$ \cite{Priede_etal07,Priede_Gerbeth09,Ruediger_etal06}, which
was also confirmed experimentally
\cite{Stefani_etal07,Stefani_etal09}. Since $\Lambda$ mainly
determines the linear dynamics of HMRI, it is natural to explore how
the nonlinear dynamics and saturation properties of HMRI depend on
this parameter, which is our main goal. This is the main reason why
we decided to follow the dynamics as a function of $\Lambda$. We
anticipate that, like in the linear theory, this parameter will be
decisive in the nonlinear outcome of the HMRI. Previous related
nonlinear studies \cite{Szklarski_Ruediger06,Szklarski07} focused
mostly on the variation of the saturated state's characteristics
with the Reynolds number at fixed Hartmann number. Besides, for
these parameters, HMRI was in the weakly nonlinear regime. In this
connection, we mention that the interaction parameter also plays an
important role in the dynamics of forced MHD turbulence with mean
field in the small-$Pm$ regime \cite{Zikanov_Thess98}.

\subsection{Numerical method}

We solve the basic Eqs. (1)-(3) using the pseudo-spectral code from
\cite{Guseva_etal15}. It is based on the Fourier expansion in the
axial $z$- and azimuthal $\phi$-directions and the finite-difference
method in the radial direction. The nonlinear terms are calculated
using the pseudospectral method and are de-aliased using the
3/2-rule. Integration in time is done with a second-order scheme
based on the implicit Crank-Nicolson method. Although the
inductionless approximation holds in the present case of small $Pm$,
the code updates the induction Eq. (2) in time in a general manner
without invoking this approximation. Further details of the code and
its validation tests on HD and MHD problems in TC flows can be found
in \cite{Guseva_etal15}.

The cylindrical flow domain in nondimensional units spans the radial
range $(R_i,R_o)=(1,2)$ divided into $N_r=100$ grid points,
azimuthal range $(0, 2\pi)$ and axial range $(0,L_z)=(0,10)$ with,
respectively, $N_{\phi}=80$ and $N_z=400$ spatial Fourier modes. The
radial grid points are placed at collocation points of Chebyshev
polynomials (despite the code is based on finite differences in $r$)
that ensures higher resolution near the cylinder surfaces. Thus,
although the HMRI is thought to be predominantly axisymmetric at
least in the linear regime, we still allow for non-axisymmetric
modes to naturally develop in the nonlinear regime by encompassing
azimuthal wavenumbers, $m$, in the range $(-N_{\phi}/2,
N_{\phi}/2)$. A resolution test we performed (not shown here) showed
that the adopted resolution is well sufficient for the present
problem, capturing a whole range of wavenumbers from dynamically
important intermediate ones, corresponding to the most unstable HMRI
modes, down to wavenumbers dominated by viscous dissipation, as
indicated by the energy spectra calculations in section 3.3 below.
The radial boundary conditions at the cylinders are standard no-slip
for the velocity perturbation, ${\bf u}|_{r=R_i,R_o}=0$, and
insulating for the magnetic field perturbation, meaning that current
does not penetrate into cylinders, that is, $\nabla \times {\bf
b}=0$ at $r<R_i$ and $r>R_o$. These radial boundary conditions are
implemented in the code (see details in
\cite{Willis_Barenghi02,Willis_Barenghi02a,Guseva_etal15}), while
periodic conditions are imposed for all variables in the axial
$z$-direction, with a period $L_z$, and in the azimuthal
$\phi$-direction. In this first effort to gain a deeper insight into
the process of the nonlinear saturation of HMRI, we vary the
$\Lambda$ number at fixed $Re$ and other parameters of the flow
(which is equivalent to varying $Ha$).

\subsection{Energy and torque}

The perturbation energy density, composed of kinetic and magnetic
parts, $e=({\bf u}^2+Ha^2Pm\cdot {\bf b}^2)/2$ (here ${\bf b}$ is
the normalized magnetic field perturbation as defined in Eqs. 5-7),
and torques at the cylinders, governing the evolution of angular
momentum $l=ru_{\phi}$ (in nondimensional units), are among the
important characteristics/diagnostics of the flow. We see that at
very small $Pm$, the magnetic energy is much smaller than the
kinetic one. The evolution equations for these quantities, taking
into account the boundary conditions, are readily derived from Eqs.
(5)-(7). As a result, we get for the total energy $E=\int edV$, to
leading order in $Rm$ \cite{Priede_etal07,Willis_Barenghi02a}
\begin{equation}
\frac{dE}{dt}=-\int u_ru_{\phi}r\frac{d\Omega}{dr}dV+D,
\end{equation}
where $D$ is the dissipation function
\[
D=-\frac{1}{Re}\int [(\nabla\times {\bf u})^2+Ha^2(\nabla \times
{\bf b})^2]dV.
\]
The first term in Eq. (8) is the Reynolds stress $u_ru_{\phi}$
multiplied by the shear, $rd\Omega/dr$, and describes energy
exchange between the basic flow and perturbations, which is
therefore due to the shear and is not affected by the magnetic
field. The dissipation function $D$ is always negative definite and
consists of two parts. The first part describes the usual viscous
dissipation of the kinetic energy, while the second part describes
the Joule losses and in general is comparable to the former. Since
the magnetic energy is negligible in this case, the Joule losses act
as an additional effective dissipation of special kind for the
kinetic energy through the kinetic-magnetic exchange term
proportional to $\Lambda$ in Eq. (5) \cite{Zikanov_Thess98}. Note
that a net contribution from nonlinear terms has canceled out in Eq.
(8) after averaging over the domain due to the boundary conditions.
Thus, only Reynolds stress, when positive, can act as a source of
perturbation energy, extracting it from the background flow due to
the shear. The Maxwell stress, $-b_rb_{\phi}$, does not play a role
because of the smallness of the induced field in the low-$Rm$
regime. The nonlinear term, not directly tapping into the flow
energy and therefore not changing the total perturbation energy,
acts only to redistribute energy among different wavenumbers. In the
quasi-steady saturated state, energy injection by the Reynolds
stress balances dissipation losses. Since $d\Omega/dr<0$, this
implies that the stress should be positive for the long-term
sustenance of this state.

For the total angular momentum, $L=\int ldV$, we get
\begin{equation}
\frac{dL}{dt}=G_i-G_o,
\end{equation}
where
\[
G_{i,o}=-\frac{R_{i,o}^3}{Re}\int_0^{2\pi}
\int_0^{L_z}\frac{d}{dr}\left(\frac{u_{\phi}}{r}\right)|_{r=R_{i,o}}d\phi
dz
\]
are the total viscous torques exerted, respectively, by the inner
and outer cylinders on the flow. These torques also characterize
angular momentum transport by the perturbations. Other -- advection
(i.e., Reynolds stress) and magnetic -- torques vanish after volume
integration due to the boundary conditions and do not contribute to
the total angular momentum evolution. In the equilibrium (laminar)
state, the torques at both cylinders have the same absolute value,
$G_{lam}=-(2\pi L_z/Re) R_i^3d\Omega/dr|_{r=R_i}=-(2\pi L_z/Re)
R_o^3d\Omega/dr|_{r=R_o}=4\pi L_z(1-\mu)a^2/Re(1-a^2)(1-a)^2>0$,
implying that the inner cylinder tries to increase the angular
momentum of the basic flow, while the outer cylinder to decrease.
These two torques are in balance, resulting in the stationary
rotational profile (4). Below we measure the torque in the perturbed
state against the laminar torque, i.e., consider the ratio
$G_{i,o}/G_{lam}$ and redefine $G_{i,o}/G_{lam} \rightarrow
G_{i,o}$. Its advantage is that it does not depend on the
normalization of the velocity.

\begin{figure}
\includegraphics[]{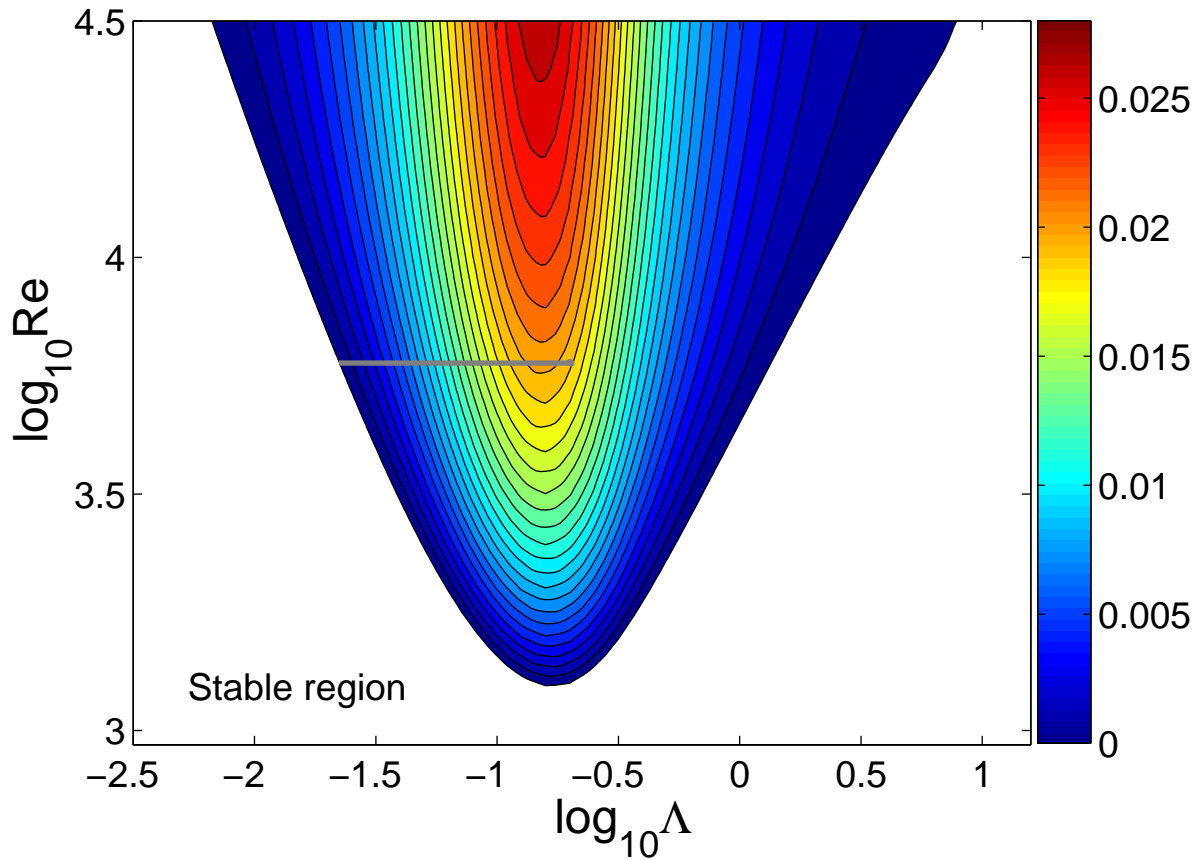}
\caption{Growth rate (in units of $\Omega_i$) of the most unstable
mode, which is axisymmetric ($m=0$), as a function of $\Lambda$ and
$Re$ (the white area is the stable region) derived from the linear
stability analysis. Note that at sufficiently high $Re\gtrsim 3000$,
the region of maximum growth is practically independent of $Re$ and
is determined mainly by $\Lambda$. The horizontal gray line
corresponds to the range of $\Lambda$ at fixed $Re=6000$ for which
our direct numerical simulations have been performed. At this value
of $Re$, HMRI sets in at $\Lambda_c=0.022$, reaches a maximum at
$\Lambda=0.15$ and then decreases again, disappearing at
$\Lambda=1.38$.}
\end{figure}
\begin{figure}
\includegraphics[]{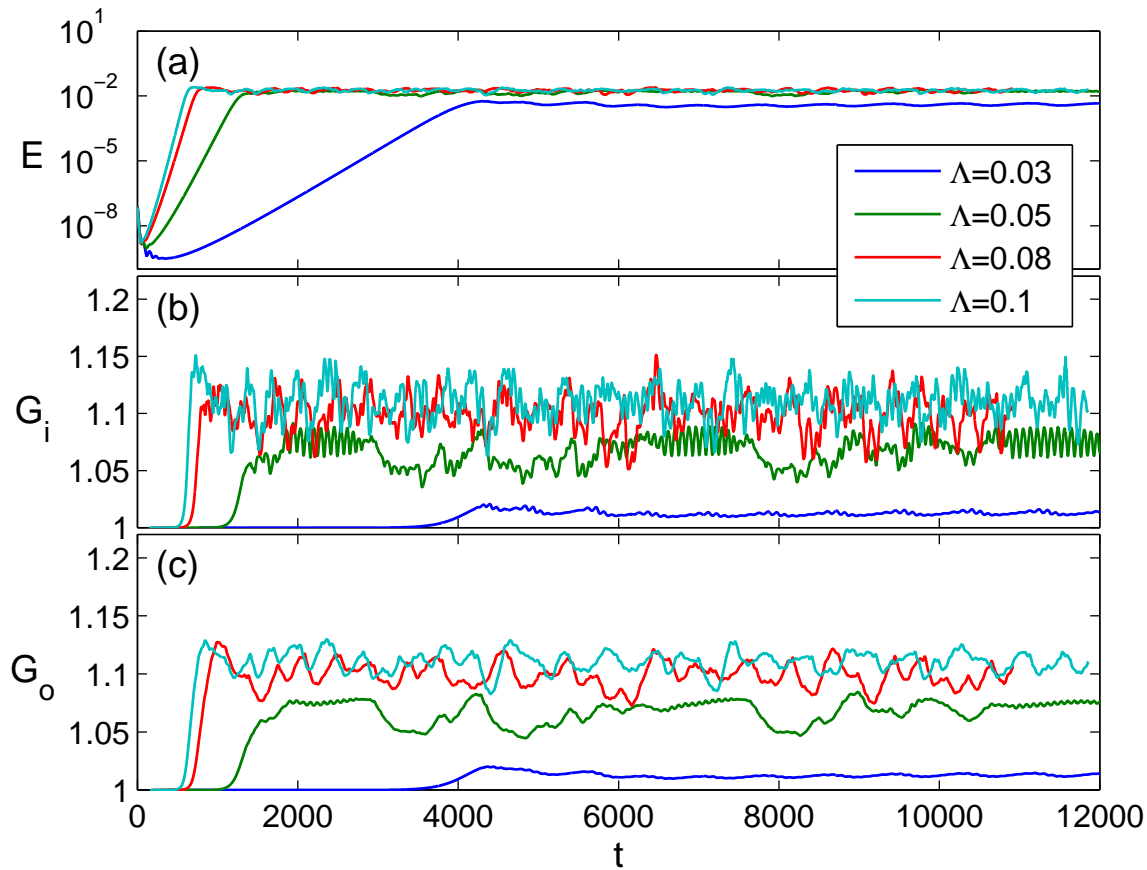}
\caption{Evolution of the volume-averaged (a) kinetic energy and
(b,c) torques at the cylinders at different $\Lambda=0.03, 0.05,
0.08, 0.1$ (i.e., $Ha=13.4, 17.3, 21.9, 24.5$). After the initial
exponential growth phase, these quantities settle down into a
sustained quasi-steady state, where they oscillate around constant
values.}
\end{figure}
\begin{figure}
\includegraphics[]{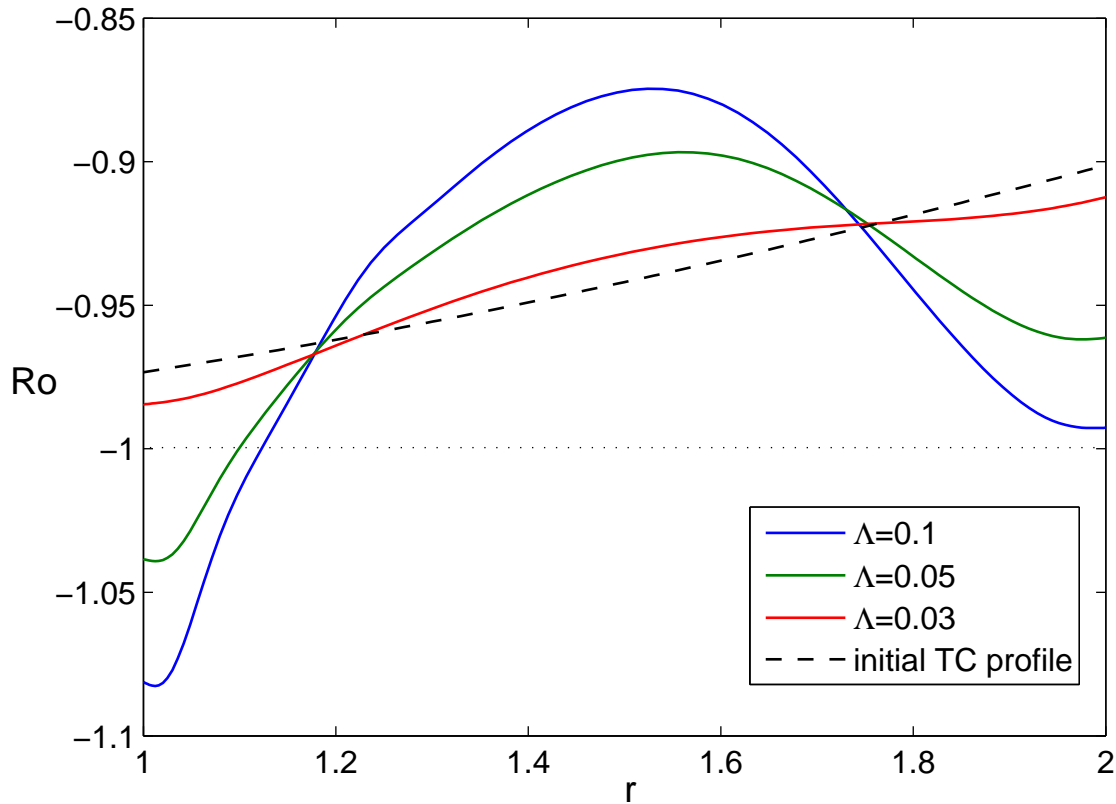}
\caption{Radial profile of the Rossby number in the saturated state
at $\Lambda=0.03, 0.05, 0.1$ calculated from the total angular
velocity, $\Omega+u_{\phi}/r$, averaged in the azimuthal $\phi$ and
axial $z$ directions. The initial radial profile of the Rossby
number corresponding to the equilibrium TC flow is shown as the
black dashed line. The Rayleigh line (dotted) corresponds to $Ro =
-1$.}
\end{figure}
\begin{figure}
\includegraphics[]{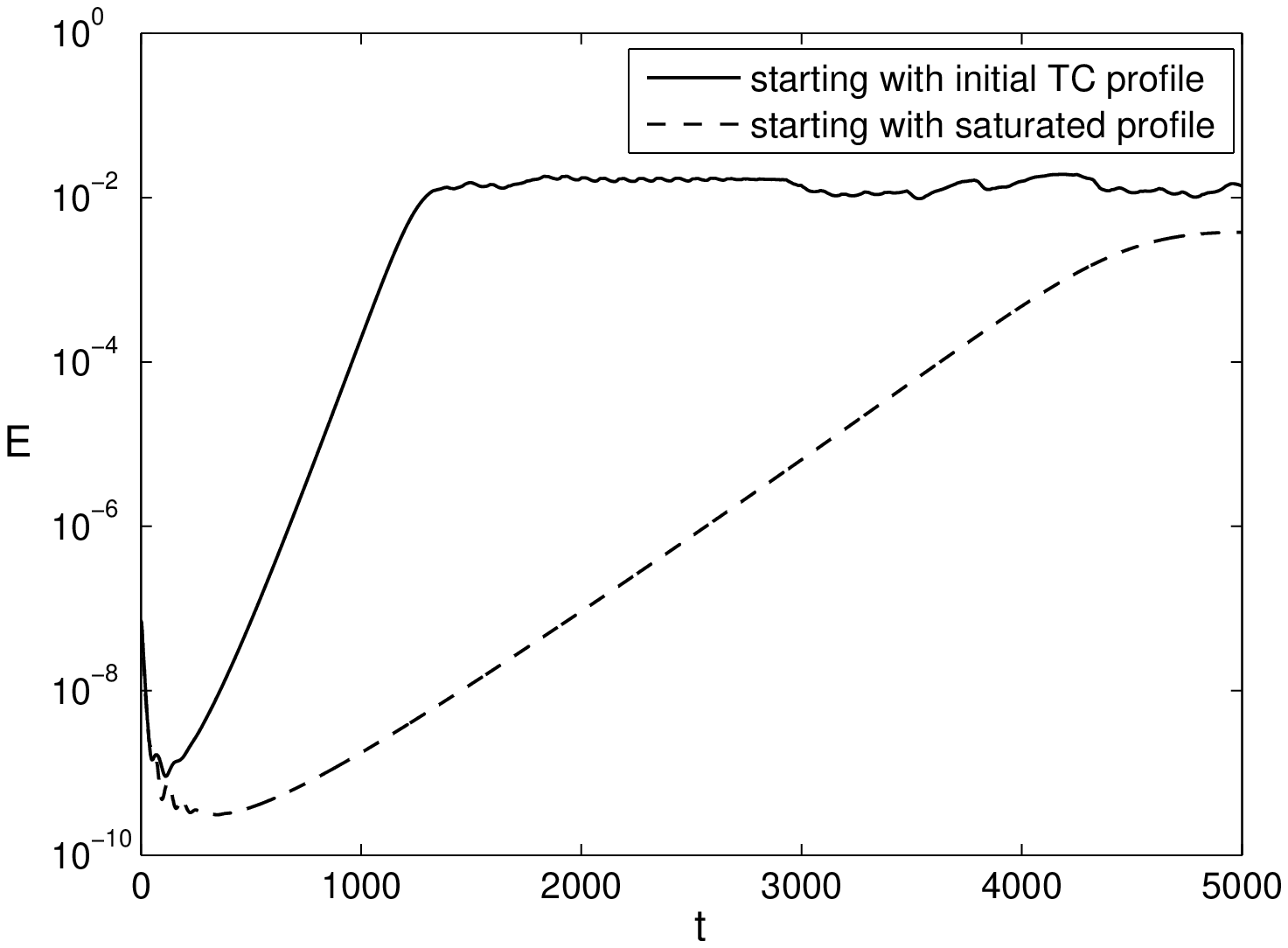}
\caption{Parallel evolution of the energy of small perturbations at
$\Lambda=0.05$. The solid line corresponds to the time-history,
starting with the initially imposed TC profile (4), (i.e., for the
same run shown in figure 2), while the dashed line to that starting
with the modified azimuthal velocity profile, which is established
in the saturated state at this $\Lambda$, having the radial
distribution of $Ro$ shown in figure 3 (green line). Calculating the
inclinations of these two linear curves, we infer that the latter
velocity profile indeed results in an 3.77 times reduced growth rate
of HMRI than the former.}
\end{figure}

\section{Results}

We move to the analysis of the HMRI, from its linear growth to
nonlinear saturation. To have an idea about the regimes of the onset
and operation of HMRI in the given TC flow configuration, in figure
1 we show the growth rate of the most unstable mode (which is
axisymmetric) as a function of $\Lambda$ and $Re$, obtained by
solving the linear stability problem with the same parameters and
boundary conditions. From this figure we see that HMRI operates in a
certain interval of $\Lambda$ provided the Reynolds number is larger
than a critical value, $Re>Re_c=1244$. The instability interval with
respect to the interaction parameter broadens with the increase of
$Re$, but converges in the inviscid limit (not shown in this
figure). It is also seen in figure 1 that at sufficiently high
$Re\gtrsim 3000$, the maximum growth of HMRI is mainly determined by
$\Lambda$ and attained at its nearly constant value. For the chosen
$Re=6000$, the instability first appears in the flow at the critical
value $\Lambda_c=0.022$, reaches a maximum at $\Lambda=0.15$ and
then decreases again, disappearing at $\Lambda=1.38$. Having
outlined the main linear features of HMRI, we can now move to
nonlinear evolution.

We initialize the simulations by imposing random noise perturbations
of the velocity and magnetic field on the equilibrium flow. The
subsequent evolution of the total energy $E$ and torques $G_i,G_o$
are shown in figure 2 at different interaction parameters, which are
all above the critical value $\Lambda_c=0.022$ when HMRI first
emerges. (We also checked that, as expected, the runs at
$\Lambda<\Lambda_c$ eventually decay, although a further study with
a variable initial amplitude of perturbations is required to see
whether there is a possibility of subcritical transition in this
case.) As we will see below, the dynamics in these cases differ
qualitatively not only in terms of the temporal evolution, but also
in its spatial appearance and spectral characteristics. After an
initial transient phase, the most unstable mode eventually emerges
and grows. The exponential growth phase lasts until the velocity
amplitude becomes large enough for the nonlinear term in Eq. (5)
(which is the dominant nonlinearity at $Rm\ll 1$) to come into play
and halt this growth. The duration of this phase depends on
$\Lambda$: if it is modestly larger than $\Lambda_c$, the saturation
takes a fraction of viscous time $O(Re)$, whereas if $\Lambda$ is
only slightly above $\Lambda_c$, saturation can take much longer,
from several to tens of viscous times. This divergence of the
saturation time at the bifurcation point (critical slowing down) is
well known from dynamical systems (e.g., \cite{Kuznetsov}). As a
result, the flow settles down into a statistically steady state
where total energy and torques oscillate around well-defined mean
values. The time variations become more irregular and the saturated
energy and torque increase with increasing the interaction
parameter. This behavior of the saturation level with $\Lambda$ will
be investigated below in more detail. Because the volume-averaged
angular momentum is also quasi-stationary in time, both torques have
nearly the same absolute values, although the inner torque is more
oscillatory at larger $\Lambda$ due to the appearance of smaller
scale turbulent eddies near the inner cylinder (see below). This
implies that the HMRI, like the SMRI, transports angular momentum
outwards, which is expected, since it also derives energy from
shear.

\subsection{Saturation mechanism}

Let us see what is a main mechanism underlying the saturation of
HMRI. During its initial exponential growth, the main nonlinear term
(advective derivative in Eq. 5) also gradually gains strength and
transfers the energy from the most unstable wavenumbers to larger
and smaller ones. In the saturated state, the linear energy
extraction rate due to the Reynolds stress associated with the
unstable wavenumbers matches the nonlinear transfer rate from these
wavenumbers to larger ones, where energy is ultimately dissipated
due to viscosity. Together with $\Lambda$, the linear growth rate of
HMRI is determined by the radial shear of the azimuthal velocity
characterized by the Rossby number, $Ro=(r/2\Omega)d\Omega/dr$
(e.g., \cite{Liu_etal06,Kirillov_etal14}), which is a function of
radius in a TC flow (for a Keplerian flow $Ro = -0.75$ everywhere).
In the exponential growth phase, a slowly increasing feedback of the
nonlinear term on the large-scale azimuthal velocity (which is
initially a TC profile given by Eq. 4) rearranges the radial
distribution of shear ($Ro$), such as to reduce it by absolute value
in the bulk of the flow. This is clearly illustrated in figure 3,
which shows the radial profile of $Ro$, calculated with the total
azimuthal velocity (i.e., initial TC flow plus the azimuthal
velocity perturbation averaged over azimuthal $\phi-$ and axial
$z$-directions) in the saturated state. It considerably deviates
from the initial profile corresponding to the TC flow and results in
the reduction of the linear growth rate of the HMRI and hence of the
energy extraction rate from the mean flow. This is confirmed by
figure 4, which shows a parallel evolution of energies in two
simulations one starting with the saturated/modified mean azimuthal
velocity profile for the run shown in figure 3 at $\Lambda=0.05$ and
the other starting with the original TC profile (4) for the same
other parameters of the flow, following only early exponential
growth phase in the linear regime. Comparing the growth rates,
derived from the inclination of ${\rm ln}(E)$ versus time, we
clearly see that the saturated profile indeed results in a reduced
exponential growth rate for the perturbation energy in the initial
linear regime, which for the chosen value of $\Lambda$ is smaller
than that in the presence of originally imposed TC flow by a factor
of 3.77. Saturation sets in when the energy gain by the linear
instability, being reduced by the nonlinearity, eventually becomes
equal to the nonlinear transfer of the energy. The saturation levels
in these two cases also differ, because, as we checked, the
corresponding established nonlinear states have different structures
-- the number of rolls in the axial direction. Thus, there is no
unique flow attractor, which would otherwise result in the same
saturation amplitudes for these two profiles. Obviously, the
deviation of the shear profile from the initial one due to the
nonlinear feedback is larger, the higher is $\Lambda$, i.e., the
more unstable is the flow. So, the saturation, where the energy gain
by the linear HMRI is in balance with the nonlinear transfers,
occurs through the modification (reduction) of the shear of the mean
rotational velocity, which in turn defines the growth rate of HMRI.
The modified shear profile in the quasi-steady state then stays
practically unchanged during the evolution.  This type of saturation
mechanism is in fact similar to that for SMRI in TC flow discussed
in
\cite{Knobloch_Julien05,Umurhan_etal07b,Ebrahimi_etal09,Clark_Oishi16a,Clark_Oishi16b}.
As for the initially imposed mean magnetic field ${\bf B}_0$, it
basically does not change in the saturated state, because, as
mentioned above, magnetic field perturbations are usually orders of
magnitude smaller the imposed field in the low-$Rm$ regime.

\begin{figure}[t]
\includegraphics[width=0.33\textwidth]{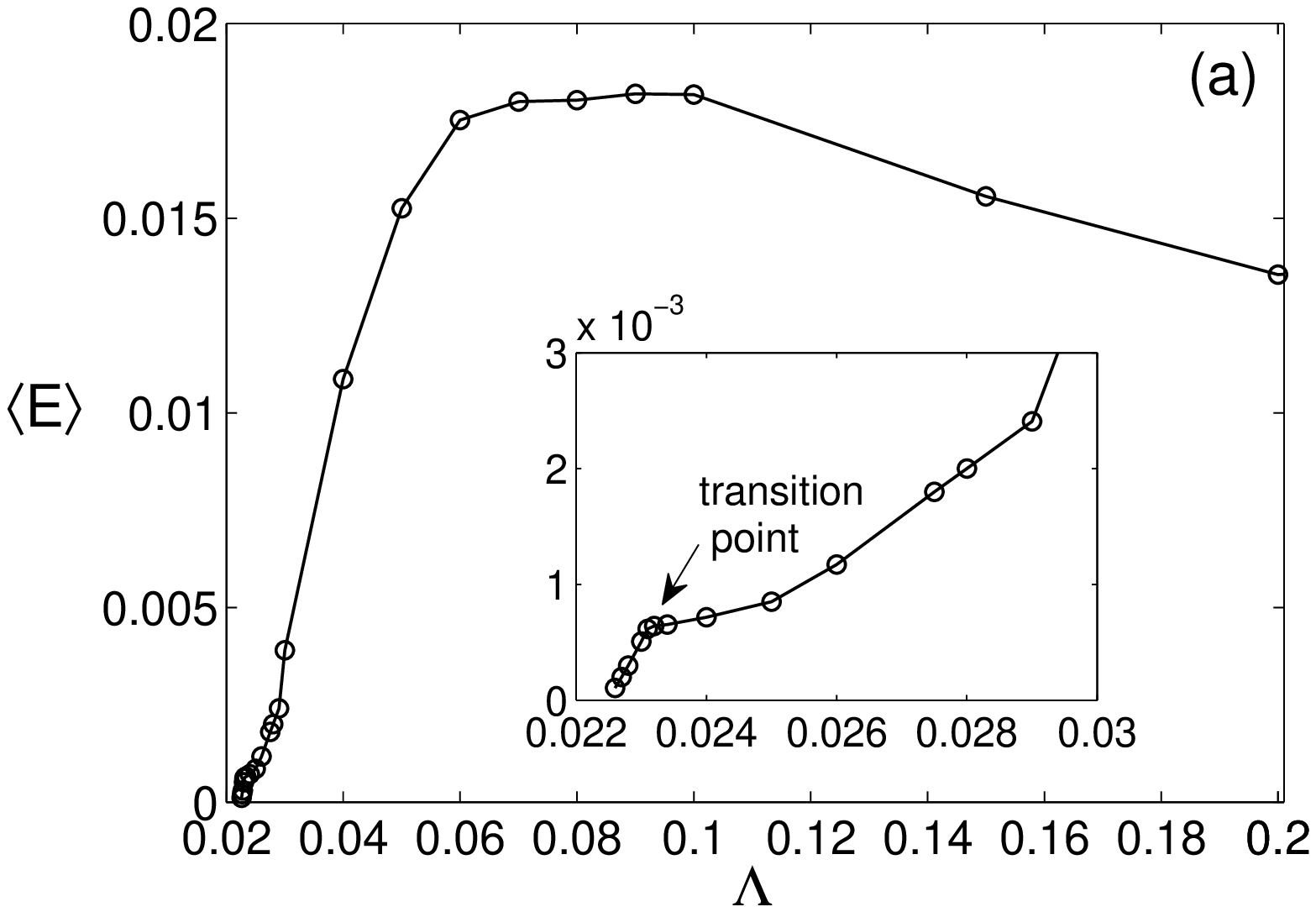}
\includegraphics[width=0.33\textwidth]{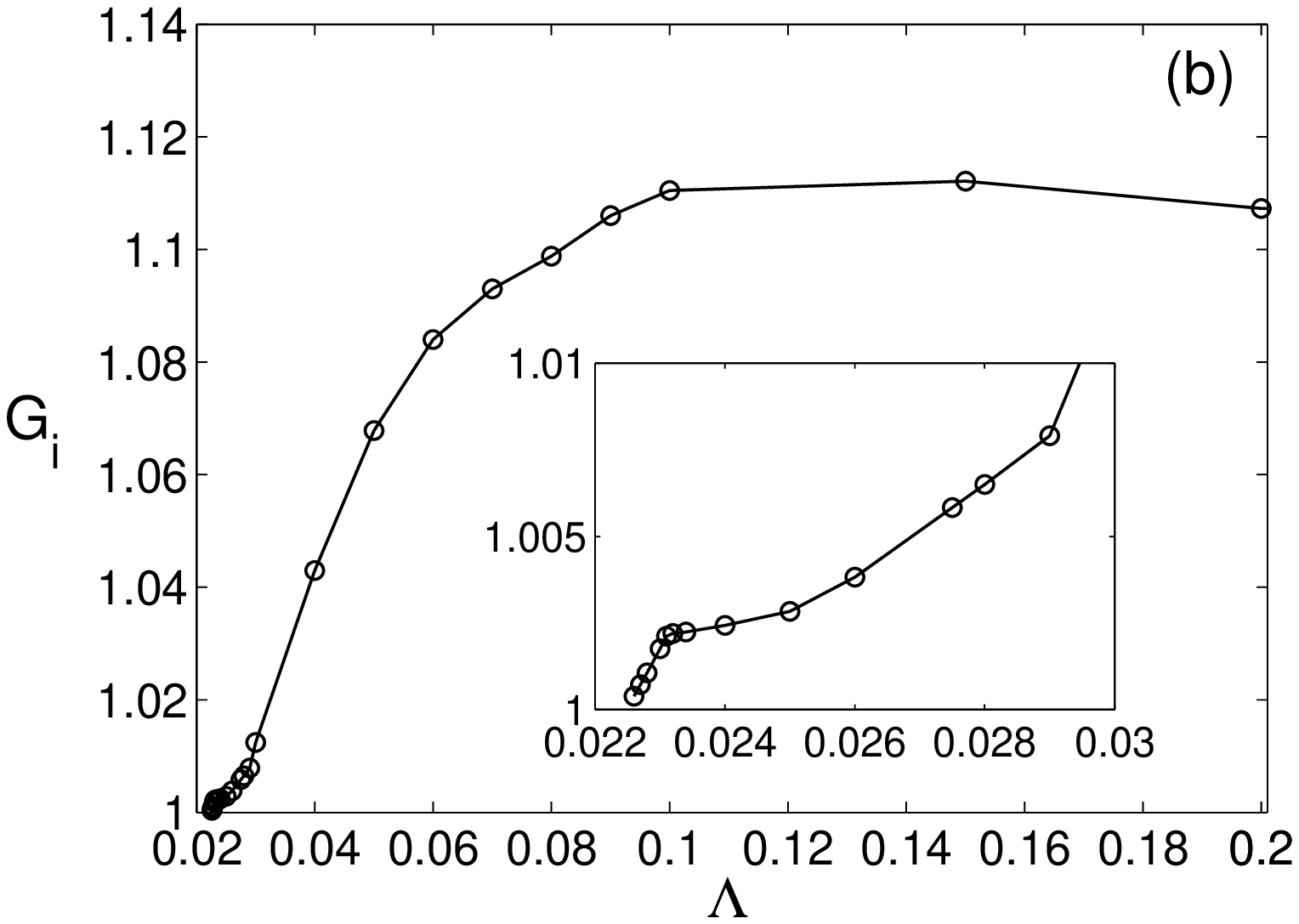}
\includegraphics[width=0.33\textwidth]{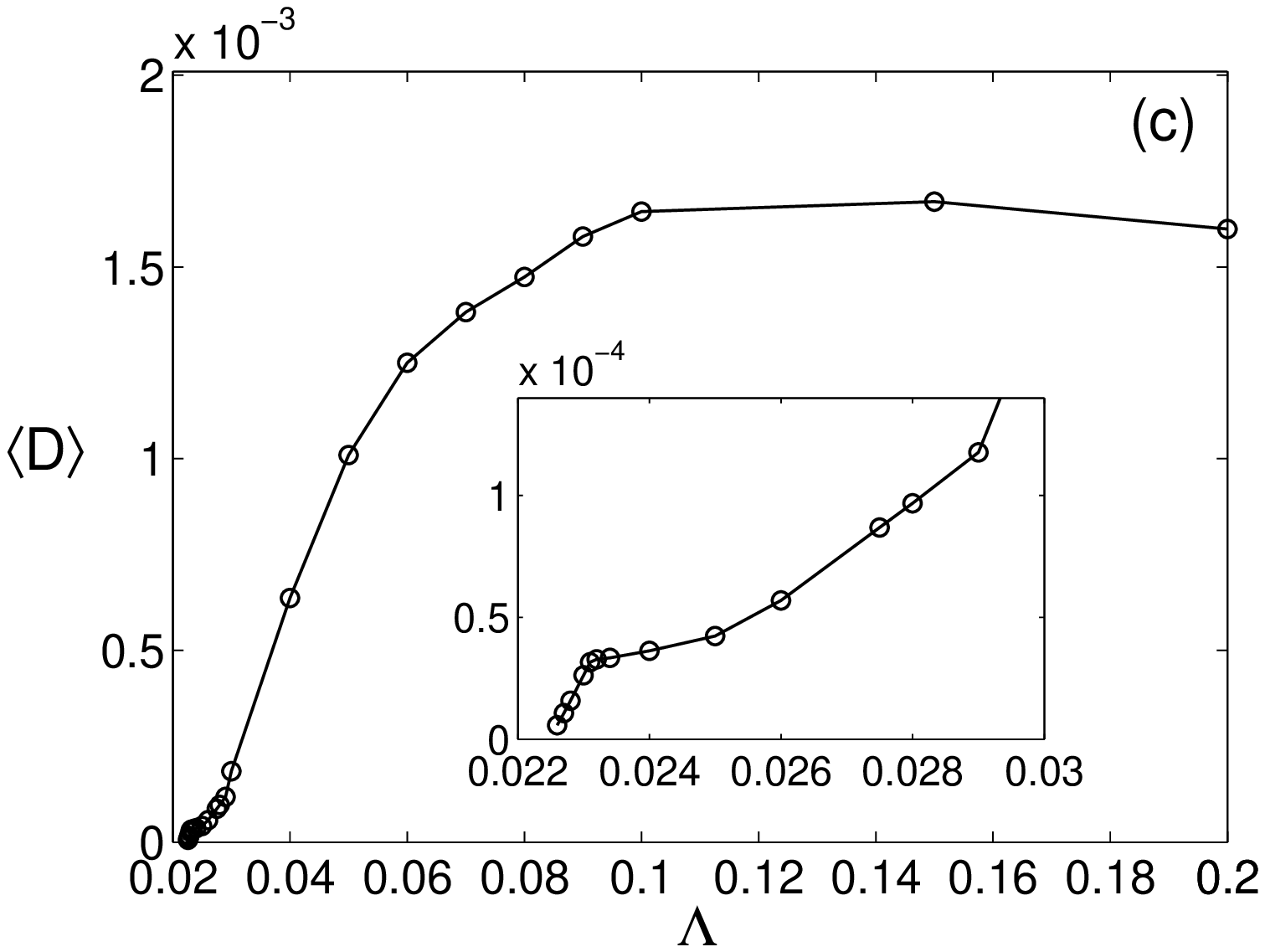}
\caption{Saturated values of the energy, $\langle E\rangle$ (a),
torque, $\langle G_i\rangle$ (b) and the dissipation, $\langle
D\rangle$ (c), as a function of $\Lambda$. Insets zoom into the
small values of $\Lambda$ corresponding to the weakly nonlinear
regime and the transition to the nonlinear chaotic regime at about
$\Lambda_{tr}=0.0232$, which is marked by a sharp inflection point
in these curves.}
\end{figure}

\subsection{Properties of the saturated state}

Figure 5 shows the time-averaged values of the energy, $\langle
E\rangle$ (a), the inner torque $\langle G_{i}\rangle$ (b) and the
dissipation rate $\langle D\rangle$ (c) in the quasi-steady
saturated state as a function of $\Lambda$. Since this state is on
average constant in time, the saturated mean values of both torques
are essentially the same, $\langle G_{i}\rangle=\langle
G_{o}\rangle$, at each $\Lambda$, as it should be (see also
\cite{Guseva_etal17}). This figure allows us to clearly see
different regimes of the nonlinear saturation of the HMRI in TC flow
and transitions between them. When the interaction parameter
increases, the instability first appears at
$\Lambda=\Lambda_c=0.022$ via classical supercritical Hopf
bifurcation \cite{Knobloch96}. For $\Lambda>\Lambda_c$, but still
near the instability threshold, the perturbations are weakly
nonlinear and the saturated energy is proportional to $\propto
\Lambda-\Lambda_c$, as expected for Hopf bifurcation. The saturated
torque and dissipation function also exhibit a similar dependence on
$\Lambda$ as the energy. As mentioned above, in this case of
$|\Lambda-\Lambda_c| << \Lambda_c$ the growth rate is small and the
saturation time is long, several tens of viscous time. A typical
spatial structure of the velocity in this regime is shown in figure
6(a), which consists of a well-organized chain of regular Taylor
vortices uniformly filling the domain and resembles the
corresponding most unstable mode from which it originated. As
evident from the isosurface plot of the axial velocity, this weakly
nonlinear state is axisymmetric. Five pairs of vortices fit in the
domain, implying that the dominant most unstable azimuthal and axial
wavenumbers are $(m,k_z)=(0,2\pi n_z/L_z)$ with $n_z=\pm 5$, which
will be confirmed by spectra calculations presented in the next
section. This picture is consistent with the pattern of the unstable
modes of the HMRI observed in previous linear stability studies
\cite{Hollerbach_Ruediger05} and nonlinear simulations in the
similar weakly nonlinear regime \cite{Szklarski_Ruediger06}. This
regime holds until about $\Lambda_{tr}=0.0232$, where an abrupt
transition to less organized and irregular (turbulent) regime takes
place, which is marked by a sharp inflection point on the curves of
the saturated values in figure 5. Beyond this point both $\langle
E\rangle$ and $\langle G_{i}\rangle$ exhibit a different behavior
with $\Lambda$ than the linear one in the weakly nonlinear regime.
When $\Lambda$ is just above $\Lambda_{tr}$, they start with a
slower increase, followed by a steeper monotonic increase starting
from $\Lambda=0.025$. The transition to the strongly nonlinear
regime is reflected in the clear change of temporal behavior of the
total energy and torques -- oscillations emerge in their evolution,
which become more and more irregular and of higher amplitude with
increasing the effect of nonlinearity as $\Lambda$ grows. This is
illustrated in figure 7, which shows the time evolution of $G_i$
before (at $\Lambda=0.023$) and after (at $\Lambda=0.026$) the
transition value $\Lambda_{tr}$ and also in figure 2 at higher
$\Lambda$. The emerging oscillations in the later case are due to a
wider spectrum of higher frequency modes (inertial waves, see below)
excited in the stronger nonlinear regime. However, already after
$\Lambda \simeq 0.08$, the saturated energy and torques do not seem
to increase anymore and the character of the time-variation are more
or less similar. In the next section, we will see that not only the
time evolution properties, but also the power spectra of the
saturated state of HMRI drastically and abruptly change when going
from the weakly nonlinear to the strongly nonlinear regime.

\begin{figure}[t]
\includegraphics[width=0.5\textwidth]{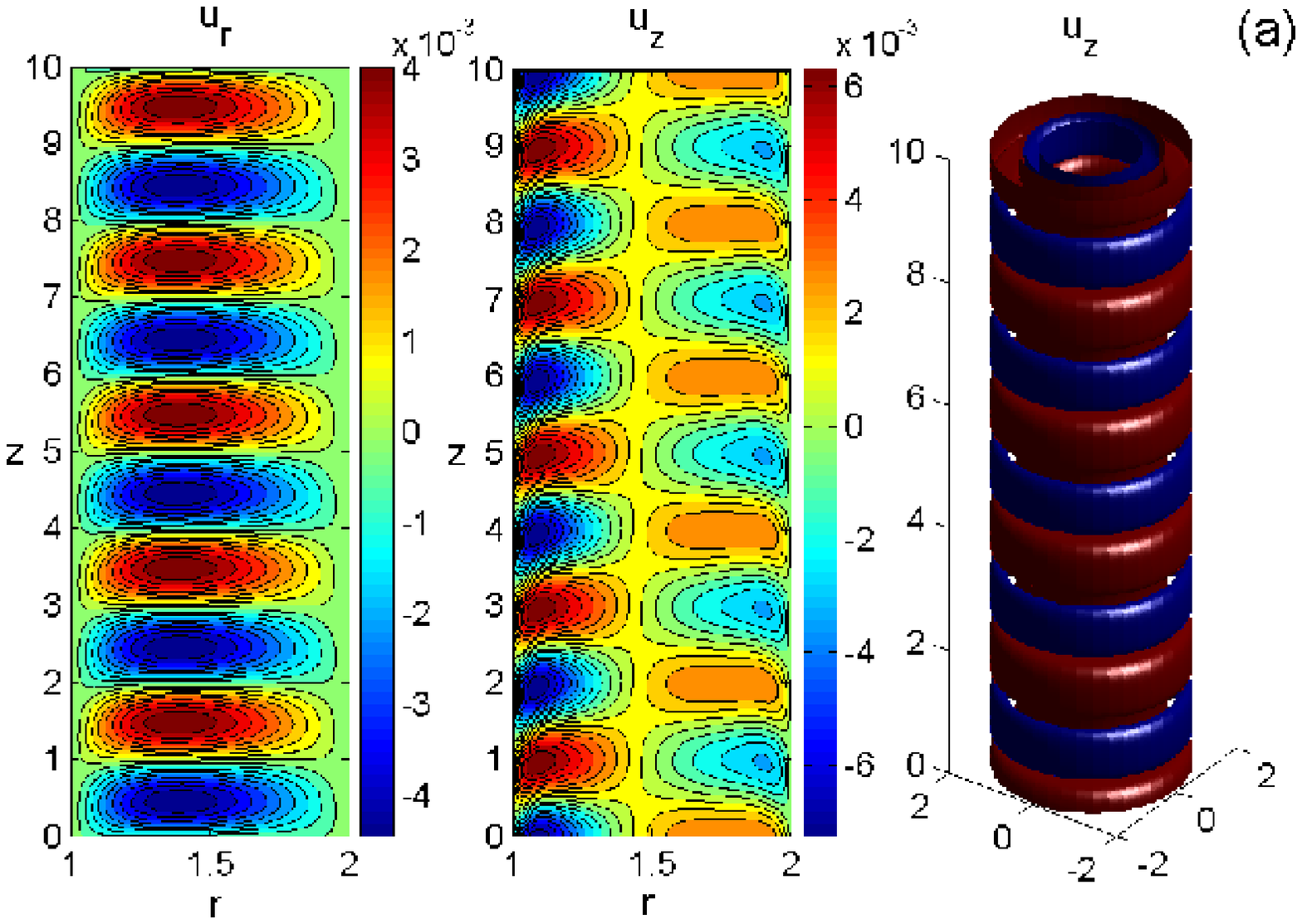}
\includegraphics[width=0.5\textwidth]{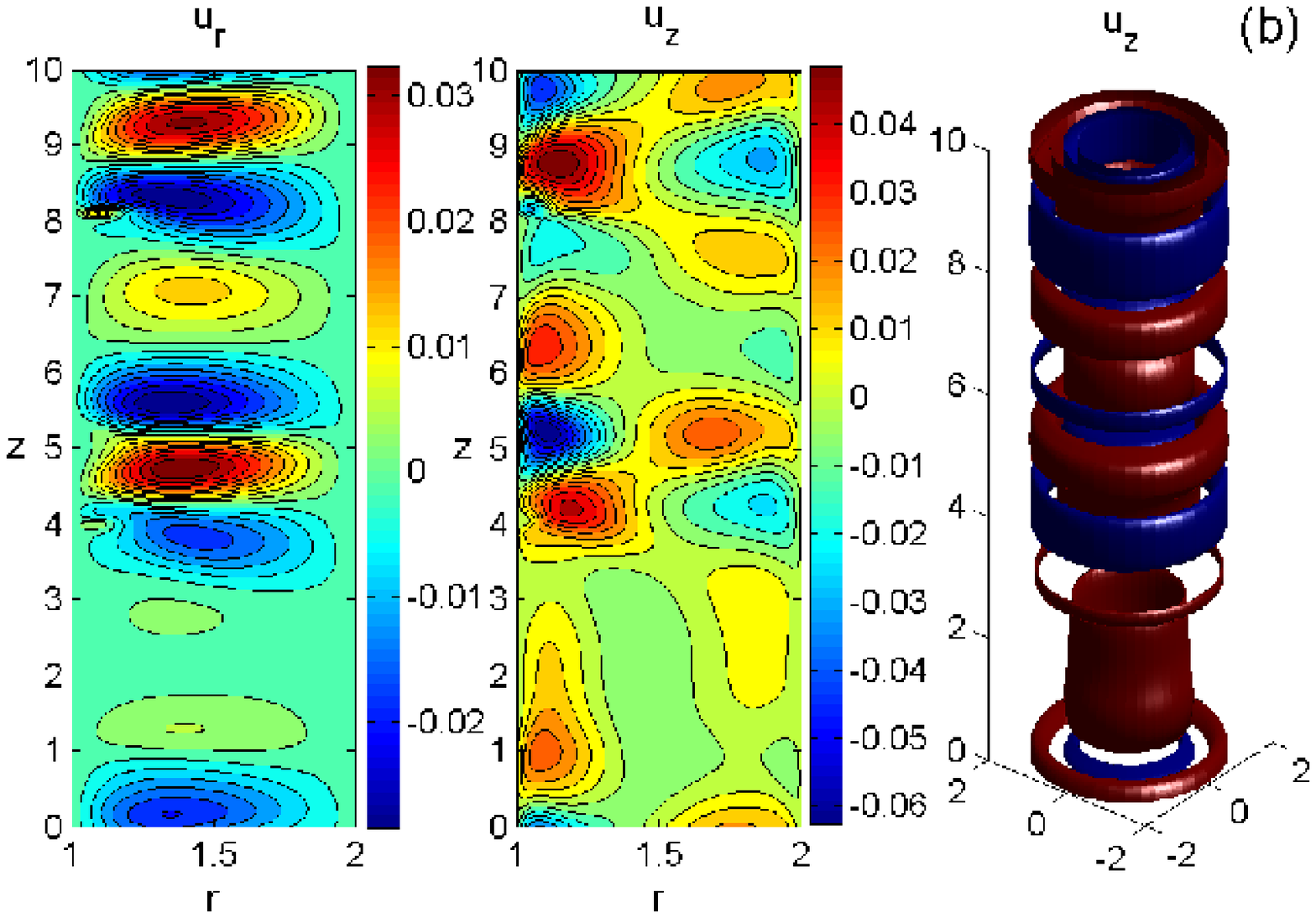}
\includegraphics[width=0.5\textwidth]{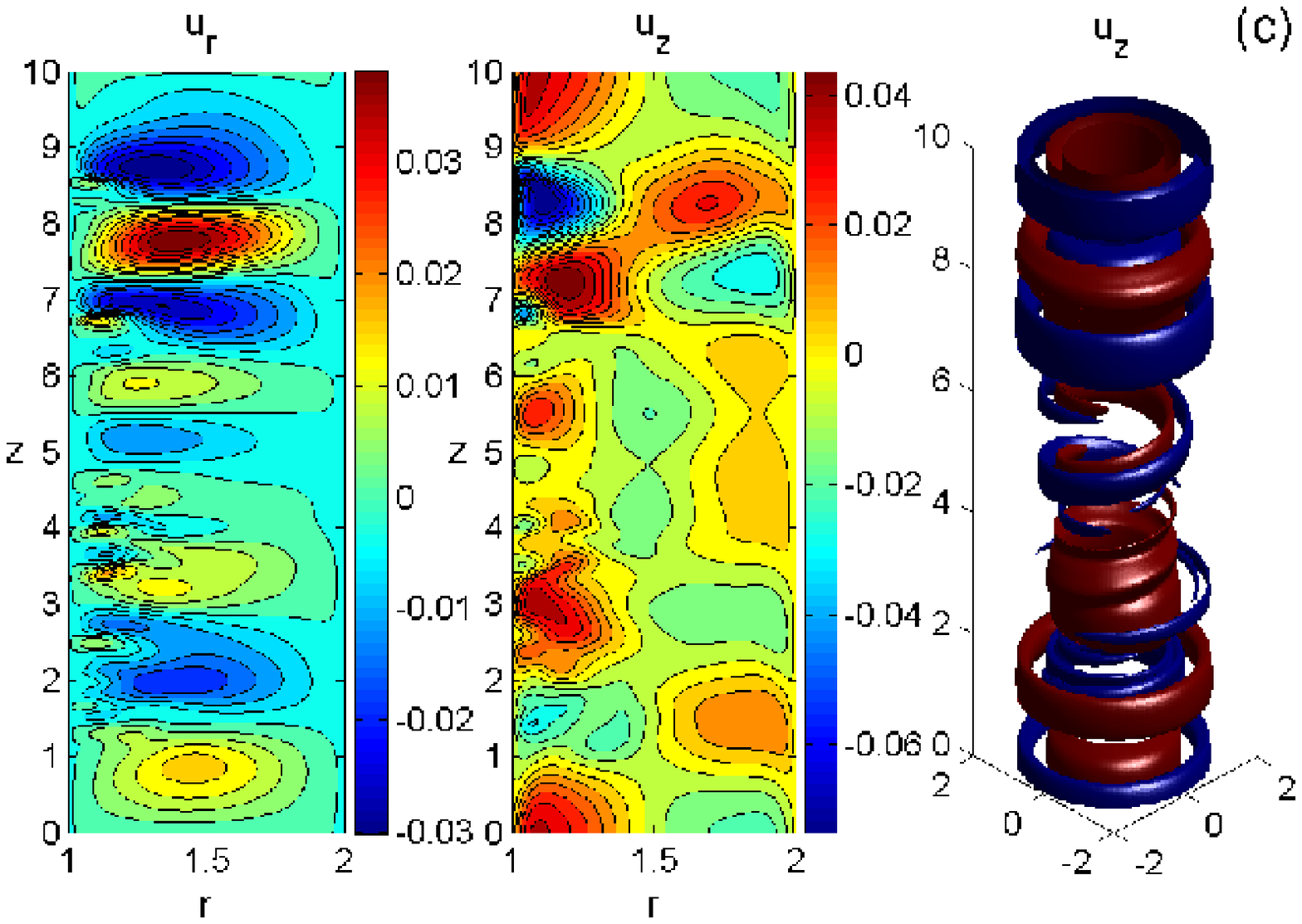}
\includegraphics[width=0.5\textwidth]{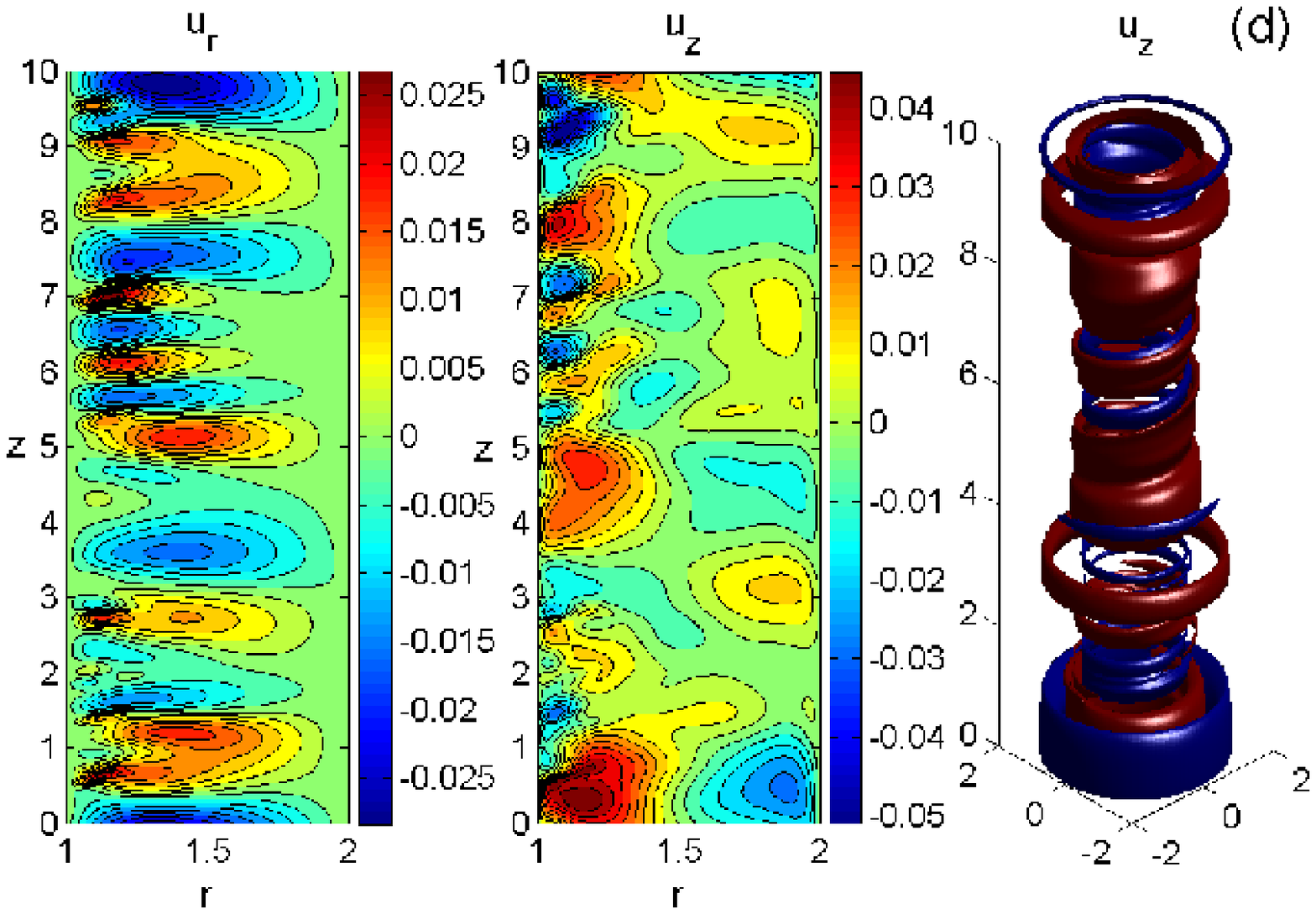}
\caption{Spatial distribution of the radial $u_r$ and axial $u_z$
velocities in the saturated state for $\Lambda=0.023 (a),0.05 (b),
0.1 (c), 0.2(d)$. Shown are the sections in $(r,z)$-plane and the
isosurfaces of the axial velocity (blue denotes negative and red
positive values). Increasing $\Lambda$, the saturated state changes
from (a) weakly nonlinear regular HMRI wave to the turbulent state
with different structures, dominated by (b) larger scale eddies or
(c),(d) by many smaller scale eddies near the inner cylinder and
larger ones near the outer cylinder. All these states are nearly
axisymmetric, although the contribution of non-axisymmetric modes
are noticeable in (c) and (d) cases and are mainly attributable to
smaller scale eddies.}
\end{figure}

\begin{figure}
\includegraphics[]{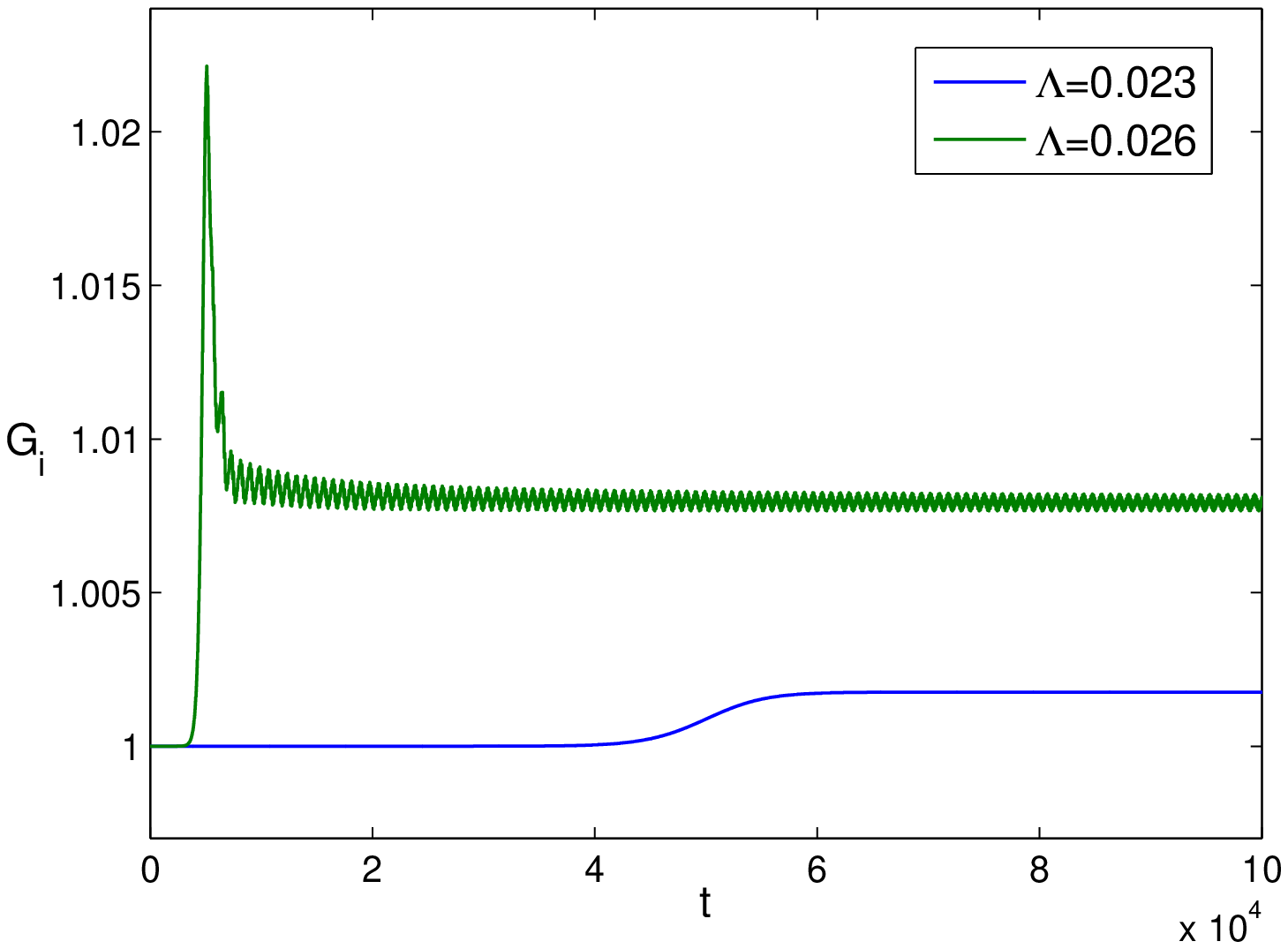}
\caption{Different character of time evolution and saturation of the
torque $G_i$ at the inner cylinder just before (at $\Lambda=0.023$)
and after (at $\Lambda=0.026$) the transition point
($\Lambda_{tr}=0.0232$). In the latter case, oscillations appear in
the saturated state and become more and more chaotic with increasing
$\Lambda$ as seen in figure 2, where larger values of $\Lambda$, far
from transition point, are taken.}
\end{figure}

Figures 6(b)-6(d) show the spatial structures of the radial $u_r$
and axial $u_z$ velocities in the saturated state of HMRI at
different $\Lambda$ after the transition point, which differ
qualitatively from those in the weakly nonlinear regime in figure
6(a). Now the eddies of different shapes and sizes have appeared in
the domain, being distributed non-uniformly along the axial
direction. With increasing $\Lambda$, more and more small-scale
vortices emerge mostly near the inner cylinder, while larger ones
remain near the outer cylinder (figures 6(c) and 6(d)). As a result,
the flow takes the form of fully developed turbulence. This is most
clearly seen in the maps of the axial velocity. These eddies change
shape in a random manner on the turnover (dynamical) time, which is
the shorter the smaller is the eddy. As a result, the torque at the
inner cylinder, $G_i$, displays faster chaotic oscillations due to
the small-scale eddies than the torque  at the outer cylinder,$G_o$,
due to larger scale eddies, as seen in figures 2(b) and 2(c). Note
also how the axisymmetry of the nonlinear state changes with
$\Lambda$ (see also figure 9 below). After the transition point
$\Lambda_{tr}$, before the emergence of smaller-scale eddies at
higher $\Lambda \gtrsim 0.1$, the nonlinear state preserves
near-axisymmetry, as evidenced by the isosurface plots of the axial
velocity, for example, in the case of $\Lambda=0.05$ in figure 6(b).
So, we observe a type of self-sustained quasi two-dimensional (2D,
i.e., dependent only on $r$ and $z$ coordinates) MHD turbulence
driven/fed by HMRI, analogous to that previously found in
hydrodynamically unstable TC flow with much stronger and purely
azimuthal background magnetic field \cite{Zikanov_etal14}. We
calculate its spectral characteristics in the next section. However,
with further increasing $\Lambda$, starting from about
$\Lambda=0.07$, the nonlinear state slightly deviates from
axisymmetry, as shown in figures 6(c) and 6(d), respectively, for
$\Lambda=0.1$ and $0.2$ (see also figure 9 below). One can discern
in these plots that the non-axisymmetry is associated mostly with
smaller-scale eddies near the inner cylinder, while the larger scale
ones near the outer cylinder stay mostly axisymmetric. Noticing that
the saturated mean azimuthal velocity profile in figure 3 develops a
Rayleigh-unstable region, $Ro < -1$, near the inner cylinder at this
larger $\Lambda$, we may conjecture that the emergence of
small-scale eddies (spatial defects) in the same radial interval is
due to a HD instability of the nonlinearly modified flow profile.
The spectral analysis presented below will give a more quantitative
characterization of the degree of the non-axisymmetry in this case.
In this regard, we would like to caution that studying the nonlinear
dynamics of HMRI at any values of the Elsasser number (i.e., for any
strength of the imposed magnetic field) with purely axisymmetric
simulations might lead to the incomplete picture of the saturation
and nonlinear dynamics of HMRI, overlooking the role of these modes.
An advantage of the present analysis, including non-axisymmetric
modes, is that it enables us to establish at which magnetic field
strength (i.e., interaction parameter) the nonlinear saturated state
of HMRI deviates from axisymmetric configuration (see below).

The transition from weakly nonlinear to strongly nonlinear and
ultimately turbulent regime for HMRI is actually analogous to the
nonlinear transition found for AMRI with increasing the Hartmann
number of the imposed purely azimuthal field at constant $Re$ by
Guseva et al. \cite{Guseva_etal15}. (Since Reynolds number is fixed,
increasing the interaction parameter corresponds to the increase of
Hartmann number.) As in our case, in the weakly nonlinear regime, a
regular pattern of vortices emerge in the saturated state via Hopf
bifurcation, except that for HMRI it is a traveling wave, as will be
evident from spectral analysis below, while for AMRI it is a
standing wave. With further increase of Hartmann number, so-called
defects appear. The spatial structure at this stage is similar to
that shown in figure 6(b) with nonuniformly distributed irregular
vortices, not far from the transition point. Subsequently, a fully
developed turbulent state arises from these defects at even higher
Hartmann numbers -- a stage which can be identified in our case with
that depicted in figures 6(c) and 6(d). It is also expected that a
similar hysteresis phenomenon mediated by an edge state dividing
weakly nonlinear and strongly nonlinear (turbulent) states, as found
by Guseva et al. for AMRI, also exists for HMRI. However, we did not
study this possibility here: the system is always followed starting
with the same random initial conditions when changing the
interaction parameter. In this case, after the exponential growth,
the flow always ends up in the chaotic state if $\Lambda$ is larger
than the transition value. Identifying edge states and the
hysteresis in the transition process for HMRI will be a subject of
future study. Such edge states are known to play an important role
in the turbulence transition in HD shear flows (see
\cite{Avila_etal13} and references therein).

\subsection{Spectral characteristics}

Further insight into the nature of the quasi-steady states of the
HMRI at different interaction parameters, which we have described so
far in physical space, can be gained by spectral analysis. It is
well known from linear theory that in the case of the background
helical magnetic field, $z$-reflection symmetry is broken
\cite{Knobloch96,Hollerbach_Ruediger05,Kirillov_etal14} and,
consequently, in the given right-handed configuration of the axial
and azimuthal fields ($\Omega B_{0\phi}B_{0z}>0$), HMRI is
characterized by inertial waves that travel downwards (opposite the
$z$-axis) \cite{Liu_etal06,Kirillov_etal14}. As a result, the
subsequent nonlinear state is composed of these downward propagating
waves. In a real experiment, both upward and downward propagating
waves are present because of the reflection from the endcaps, but
the one that is unstable tends to be dominant \cite{Stefani_etal09}.
Below we analyse the axial and azimuthal spectra of these waves. To
this end, we decompose each velocity component in the azimuthal and
axial directions (in which they are periodic),
\[
\bar{u}_i(r,m,k_z)=\frac{1}{2\pi L_z}\int_0^{2\pi} \int_0^{L_z}
u_i(r,\phi,z){\rm e}^{-{\rm i}m\phi-{\rm i}k_zz}d\phi dz,
\]
where the index $i=r,\phi,z$ and $m=0,1,2,...$ and $k_z=2\pi
n_z/L_z$ with $n_z=0,\pm 1, \pm 2,...$, are the azimuthal and axial
wavenumbers, respectively. We also define radially integrated
spectral energy density,
\[
{\cal E}(m,k_z)=\pi L_z\int_{R_i}^{R_0}
(|\bar{u}_r|^2+|\bar{u}_{\phi}|^2+|\bar{u}_z|^2)rdr,
\]
as well as the azimuthal, $\hat{\cal E}_m$, and axial, $\hat{\cal
E}$, spectral energy densities via
\[
\hat{\cal E}_m=\sum_{k_z}{\cal E}(m,k_z), ~~~\hat{\cal
E}=\sum_{m}{\cal E}(m,k_z),
\]
so that their sum over wavenumbers is equal to the total energy,
$E=\sum_{m,k_z} {\cal E}(m,k_z)=\sum_{m}\hat{\cal
E}_m=\sum_{k_z}\hat{\cal E}$. These energy spectra are among the
main characteristics/diagnostics in the nonlinear (turbulent) regime
in flows. In particular, the azimuthal spectrum characterizes the
total contribution of different $m$ in the dynamics and hence can
serve as a measure of the non-axisymmetry, while the axial $k_z$
spectra provide information on the energy-injection due to
axisymmetric HMRI modes (see e.g., \cite{Zikanov_etal14}). We use
these spectra below to characterize the dynamics in the quasi-steady
state. This state, however, exhibits, as we have seen above,
irregular oscillations, making the spectra noisy. To avoid this, we
also average $\hat{\cal E}_m$ and $\hat{\cal E}$ in time over the
whole duration of the quasi-steady state in the simulations.

\begin{figure}
\includegraphics[]{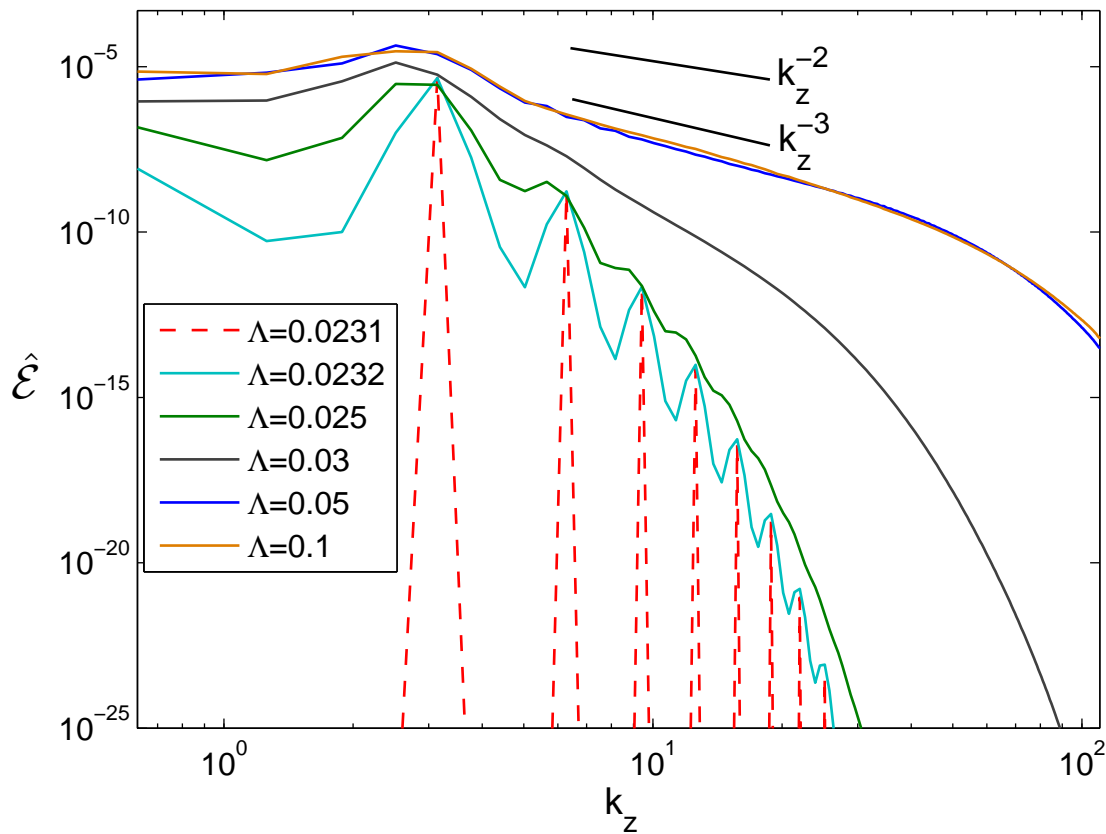}
\caption{Time-averaged energy spectra $\hat{\cal E}$ vs $k_z$ in the
quasi-steady saturated state at different $\Lambda$. For reference,
we also show the 2D HD turbulence spectrum $k_z^{-3}$ as well as a
flatter $k_z^{-2}$ spectrum of rotating HD turbulence. At small
$\Lambda$, before the transition value $\Lambda_{tr}=0.0232$, the
spectrum is discrete (``spiky'' dashed lines), dominated by the most
unstable mode and its multiple wavenumbers, while for $\Lambda$
beyond the transition point, in the strongly nonlinear (turbulent)
regime, the spectrum (solid lines) becomes continuous with all axial
wavenumbers excited; at $\Lambda \gtrsim 0.05$ it exhibits a scaling
behavior at intermediate wavenumbers $5\lesssim k_z\lesssim 20$,
which is close to $k_z^{-3}$.}
\end{figure}

Figure 8 shows the time-averaged axial energy spectrum $\hat{\cal
E}$ in the saturated state at different $\Lambda$. It is seen that
qualitatively different spectra correspond to weakly and strongly
nonlinear saturation regimes, and the rearrangement from the former
to the latter type, like the saturated values of energy and torque
(figure 5), occurs suddenly near the transition point
$\Lambda_{tr}=0.0232$. At small $\Lambda\gtrsim \Lambda_c$, the
linearly most unstable axial wavenumber of HMRI, $k_z=2\pi
n_z/L_z=3.14$ with the mode number $n_z=5$, is dominant also in the
saturated state and carries most of the power. In physical space,
this corresponds to five pair of vortices, as shown in figure 6(a).
The modes with wavenumbers, which are multiples of the most unstable
axial wavenumber, are excited due to weak nonlinear self-interaction
of the dominant mode in the saturation process. These subharmonics
(separate spikes in figure 8) have orders of magnitude smaller power
compared to the dominant one, whereas at all other $k_z$ the
spectral power is essentially zero. Remarkably, the energy spectrum
changes qualitatively when the interaction parameter approaches a
transition value $\Lambda_{tr}$, where the effect of nonlinearity is
more appreciable. As a result, power in other wavenumbers, lying
between the multiple wavenumbers, abruptly increases, i.e., the
spectrum is quickly filled and becomes continuous, although the
multiple harmonics of the most unstable wavenumber still have larger
power than other ones. But already at $\Lambda=0.025$, slightly
larger than the transition point, the spectrum takes a smoother
shape, continuously well populated at all wavenumbers, as it is
characteristic of a turbulent state. With further increasing
$\Lambda$, this spectrum moves upwards and converges starting from
about $\Lambda=0.05$. Now, the maximum comes again at the wavenumber
of the most unstable HMRI mode, $k_z\sim 2-3$, which mainly
determines the energy injection from the basic flow into turbulent
fluctuations. As it is seen in figure 8, at intermediate wavenumbers
in the range $5\lesssim k_z\lesssim 20$, this converged spectrum
displays the power-law dependence close to $k_z^{-3}$ typical for 2D
HD turbulence \cite{Kraichnan67}, as opposed to the Kolmogorov
spectrum, $k^{-5/3}$, for three-dimensional turbulence or to the
$k^{-2}$ spectrum for rotating HD turbulence \cite{Mueller07}. Then,
at $k_z\gtrsim 20$ the spectrum decays faster because of viscous
dissipation. The appearance of the quasi-2D turbulence in physical
space, corresponding to these spectra, we have already seen in
figure 6. The energy spectrum with a similar power-law dependence
were also reported for 2D MHD turbulence in TC flow with a large
purely azimuthal magnetic field in the highly resistive
(inductionless) limit, $Rm \ll 1$, in the above-mentioned work
\cite{Zikanov_etal14}. In that case, however, the turbulence is due
to HD instability and forced to be 2D by the imposed azimuthal
field. By contrast, in the present case, where the TC flow is
hydrodynamically stable, the observed turbulence is of magnetic
origin, triggered and energetically supplied by HMRI, and hence
quasi-2D from the outset. This implies that the turbulence we
observe here cannot be described within standard Kolmogorov
phenomenology, which anyway is inapplicable in the case of high
resistivity, because the Joule dissipation in this regime is
anisotropic and acts at all scales in the flow (see e.g.,
\cite{Zikanov_Thess98}). In this situation, it is not possible
generally to define an inertial range in a classical sense, where
only nonlinear term $({\bf u}\cdot \nabla){\bf u}$ in Eq. (5)
operates and transfers energy towards large wavenumbers. The
presence of the power-law interval in the energy spectrum different
from Kolmogorov one that has been found here and in
\cite{Zikanov_etal14} is a confirmation of that. The similarity of
the spectra and the associated power-law indices, despite different
driving mechanisms of the turbulence in our case and in
\cite{Zikanov_etal14}, indicates that these spectral features could
be generic to quasi-2D MHD turbulence, which can occur in a
magnetized resistive TC flow.

\begin{figure}
\includegraphics[width=0.5\textwidth]{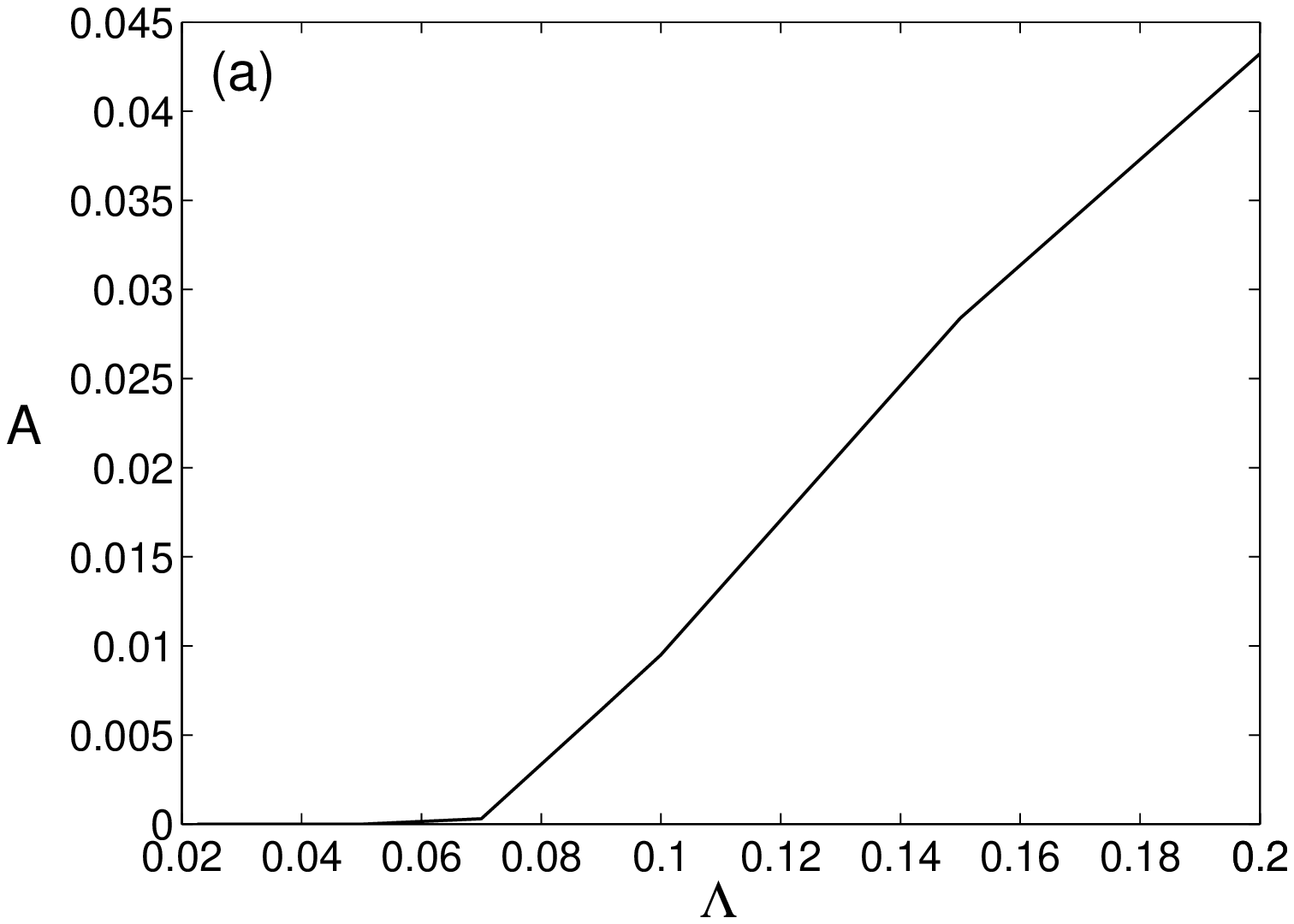}
\includegraphics[width=0.5\textwidth]{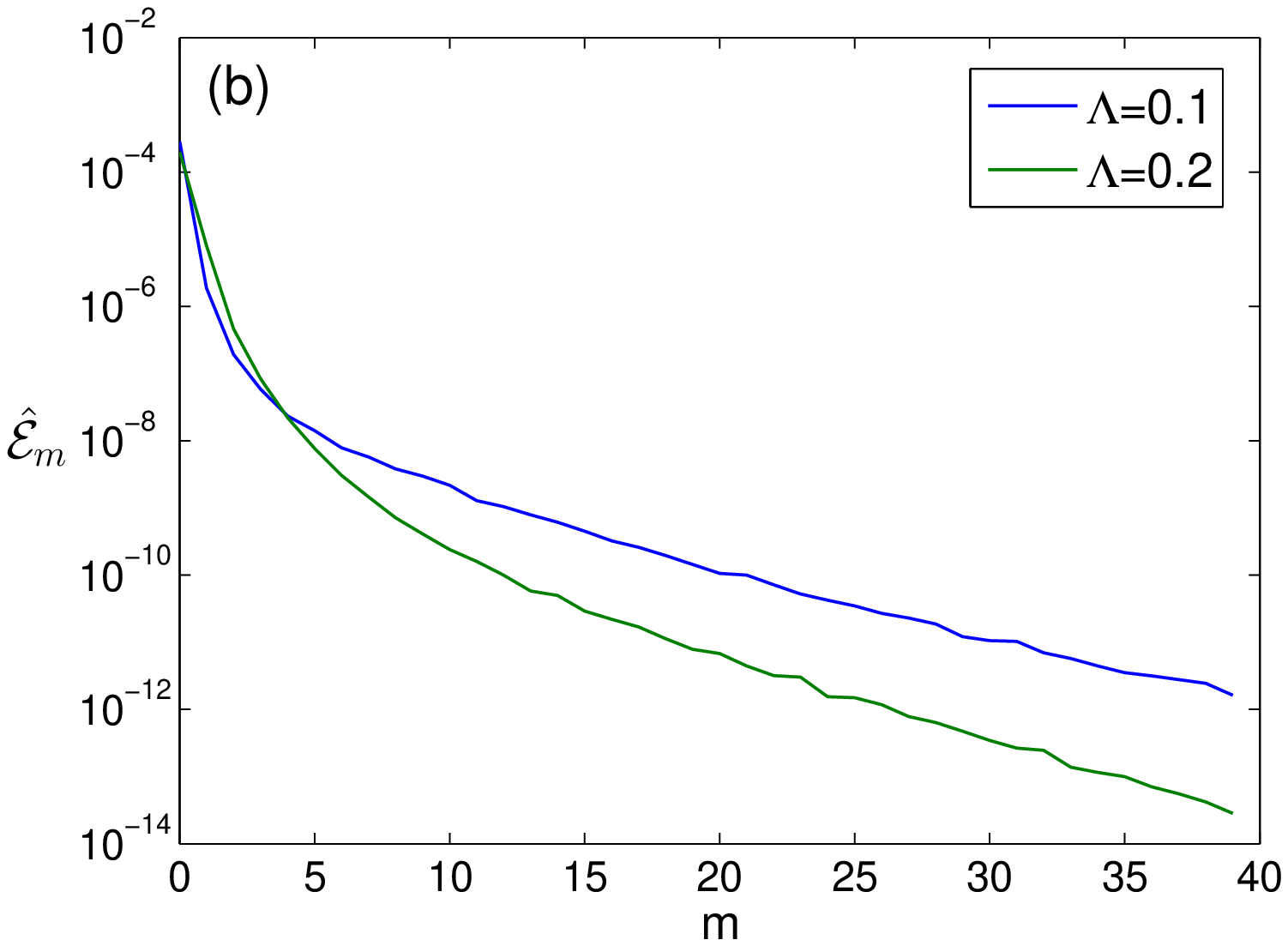}
\caption{The parameter $A$, characterizing the role of
non-axiymmetric $(m\neq 0)$ modes relative to the axisymmetric
($m=0$) ones as a function of $\Lambda$ (a) and the azimuthal
spectrum of the energy, $\hat{\cal E}_m$, at $\Lambda=0.1, 0.2$
(b).}
\end{figure}

As we have found above, the saturated state remains nearly
axisymmetric, despite non-axisymmetric modes emerging with
increasing $\Lambda$. To quantify the role of the latter modes, in
figure 9 we plot the ratio of the sum of azimuthal energies
$\hat{\cal E}_m$ of all the non-axisymmetric modes with $m\geq 1$ to
the energy of the axisymmetric modes with $m=0$, measured by the
parameter $A=\sum_{m\geq 1}\hat{\cal E}_m/\hat{\cal E}_0$, as a
function of $\Lambda$ (a) as well as the typical azimuthal energy
spectrum in the strongly nonlinear state (b). From this figure, it
is seen that non-axisymmetric modes are essentially absent at small
$\Lambda$; they start to emerge from about $\Lambda=0.07$, after
which the total energy of all non-axisymmetric modes relative to
that of axisymmetric ones increases with $\Lambda$. However, the
former still remains much smaller than the latter, particularly in
the range $0.07\leq \Lambda \leq 0.2$, where the saturation level
reaches a maximum (figure 5). This is also confirmed by the
azimuthal spectrum of the energy, $\hat{\cal E}_m$, in figure 9(b).
Since non-axisymmetric modes are present only at $\Lambda\gtrsim
0.07$, we show the azimuthal spectrum at typical values
$\Lambda=0.1$ and 0.2, corresponding to the strongly nonlinear
(turbulent) regime. It reaches a maximum at $m=0$ and rapidly
decreases with $m$. The energy of the first non-axisymmetric modes
with $m=1$ is already by more than an order of magnitude smaller
than that of the axisymmetric ones, $\hat{\cal E}_1/\hat{\cal
E}_0$=0.007 and 0.04, respectively, at $\Lambda=0.1$ and 0.2, and
the energy of higher non-axisymmetric modes is even lower.

\section{Summary and conclusion}

In this paper, we investigated the development of HMRI in an
infinite/periodic TC flow domain at very small magnetic Prandtl
numbers by following its evolution from the linear growth phase to
nonlinear saturation using direct numerical simulations. To focus on
the basic dynamics of HMRI arising from the combined effect of
differential rotation and helical magnetic field, we simplified the
analysis by ignoring the effects of endcaps that would induce Ekman
pumping. We analyzed the dynamics with respect to the interaction
parameter, or Elsasser number, $\Lambda$, which is known to be a
central parameter determining the linear evolution of HMRI. As
distinct from previous nonlinear analyzes of HMRI, which focused
only on axisymmetric modes, an advantage of our study is that by
allowing for non-axisymmetric modes in the simulations, it enabled
us to establish for which interaction parameters the role of the
latter modes becomes important. We confirmed that in the nonlinear
regime, just as SMRI, HMRI transports angular momentum outward. We
demonstrated that different regimes of nonlinear saturation are
realized when the interaction parameter changes. At smaller values
of this parameter, just above the stability threshold, the HMRI
saturates in the weakly nonlinear regime and the corresponding
spatial structure consists of well-organized regular vortices, which
are axisymmetric. In this case, the most unstable mode of
instability dominates, which is also supported by our spectral
analysis. However, with increasing the interaction parameter, at a
certain value an abrupt transition to strongly nonlinear (turbulent)
state takes place, marked by different, irregularly oscillating
behavior of the energy and stress. The saturated values of the
energy and stresses as a function of $\Lambda$ exhibit a sharp
inflection point, corresponding to this transition, and then
increase with until about $\Lambda=0.1$ and then slowly decrease
(figure 5), since the effectiveness of the HMRI itself is decreasing
at higher magnetic fields. The spatial structure of the nonlinear
state is predominantly axisymmetric and represents a type of
self-sustained quasi-2D (in $(r,z)$-plane) MHD turbulence due to
HMRI. Generally, classical 2D MHD turbulence is decaying and
requires external forcing for long-term maintenance
\cite{Biskamp03}. So, here we demonstrated the existence of
self-sustained quasi-2D MHD turbulence in a TC flow at high
resistivity, or small magnetic Prandtl numbers caused and supplied
by HMRI at the expense of the flow energy. We also calculated the
energy spectrum as a function of axial wavenumber and found that it
increases with $\Lambda$, but converges from about $\Lambda=0.05$,
and at intermediate axial wavenumbers $k_z$ exhibits a power-low
dependence close to $k_z^{-3}$, as typical for 2D HD turbulence
\cite{Kraichnan67}. It would be interesting to probe in future
studies whether this quasi-2D turbulent state persists as the
Reynolds number increases or a transition to three-dimensional
turbulence occurs.

In both weakly and strongly nonlinear regimes, the saturation
mechanism appears to be general, consisting in the modification, or
reduction of the radial shear profile of the mean azimuthal velocity
in the bulk of the flow, which, in turn, results in the reduction of
the exponential growth rate of HMRI to match energy transfer rate
due to the nonlinear (advection) term. Such a nonlinear saturation
mechanism was already discussed for SMRI in a TC flow
\cite{Knobloch_Julien05,Umurhan_etal07a,Ebrahimi_etal09,Clark_Oishi16a,Clark_Oishi16b}.

The present analysis, although simplified by excluding the effects
of endcaps, is a first step towards understanding the experimental
manifestation of HMRI, where it is already in the saturated
nonlinear regime. The insight gained from this study will be a
stepping stone for future numerical studies incorporating endcaps
and conducing boundaries, where the situation is complicated by
Ekman circulations, penetrating from the endcaps in the bulk of the
flow, and for subsequent comparison with planned experiments on HMRI
within the DRESDYN project at HZDR. Nevertheless, as demonstrated in
\cite{Guseva_etal15} on the example of AMRI, the results obtained
with periodic boundary conditions along the axial direction are
still useful to interpret the experiments. In the future, we plan to
do analogous studies of HMRI in a TC flow with endcaps and compare
with the present results. To date, such realistic three-dimensional
simulations of HMRI, which are necessary for understanding the
related experimental results, have not been undertaken yet because
of high computational cost. Although previous early simulations
explored the nonlinear development of HMRI in TC flow with endcaps,
they considered a narrow range of parameters and/or were
axisymmetric by design
\cite{Szklarski_Ruediger06,Liu_etal07,Szklarski07}, not capturing
different regimes of the nonlinear saturation.

Another interesting venue of research extending the present analysis
is to explore the nonlinear dynamics of HMRI and associated angular
momentum transport for quasi-Keplerian rotation relevant to
protoplanetary disks, whose dense and cold interiors are too
resistive for SMRI to operate. The nonlinear development of HMRI has
not been explored also for positive shear profiles relevant to the
near-equator strip of the solar tachocline. These topics are of a
current research interest and have been studied so far only in the
linear regime
\cite{Kirillov_Stefani13,Kirillov_etal14,Stefani_Kirillov15}.

\ack

This work was supported by the Alexander von Humboldt Foundation and
by the German Helmholtz Association in frame of the Helmholtz
Alliance LIMTECH. Computations were done on the high-performance
Linux cluster Hydra at the Helmholtz-Zentrum Dresden-Rossendorf. We
would like to thank R. Hollerbach for sharing and assisting with his
code and valuable discussions on HMRI at the initial stage of this
work. We also thank O. Zikanov for useful discussions on low-$Rm$
MHD turbulence.

\section*{References}
\bibliographystyle{iopart-num}
\bibliography{biblio}
\end{document}